\documentclass{JHEP3}
\usepackage{epsfig}

\newcommand{\be}{\begin{equation}}
\newcommand{\ee}{\end{equation}}
\newcommand{\bea}{\begin{eqnarray}}
\newcommand{\eea}{\end{eqnarray}}
\newcommand{\Tr}{{\rm Tr}}
\newcommand{\ub}{\overline}
\newcommand{\tr}{{\rm tr\ }}
\newcommand{\flfn}[1]{{\left\lfloor#1\right\rfloor}}
\newcommand{\clfn}[1]{{\left\lceil#1\right\rceil}}
\newcommand{\Feyn}[1]{#1\kern-0.7em/\kern 0.2em}

\title{Tata lectures on overlap fermions~\footnote{Lectures given at
  the {\bf Asian School on Lattice Field Theory 2011 (an ICTS program).}}}
\author{R. Narayanan\\
Department of Physics, Florida International University, Miami,
FL 33199.
\\E-mail: \email{rajamani.narayanan@fiu.edu}}

\abstract{
Overlap formalism deals with the construction of chiral gauge
theories on the lattice. These set of lectures provide a pedagogical
introduction to the subject with emphasis on chiral anomalies
and gauge field topology. Subtleties associated with the generating
functional for gauge theories coupled to chiral fermions are discussed.
}


\preprint{}

\begin{document}

\section{Introduction}

Chiral symmetry, chiral anomalies and gauge field
topology are important in our understanding of theories that
describe fundamental interactions.
Since the focus of these lectures is on the lattice formulation of
chiral gauge theories,
we will focus on continuum Euclidean field theory and use the
path integral representation.
It is sufficient to
consider two dimensional chiral fermions
and couple them to an abelian gauge field to understand all the
relevant concepts. We will therefore focus only on such theories.
The extension to any other
chiral gauge theory will be straight forward.

The full path integral will be broken up into two pieces:
\be 
Z = \int [dA] e^{-S_g(A)} \int [d\psi][d\ub\psi] 
e^{S_f(\ub\psi,\psi,A)}
= \int [dA] e^{-S_g(A) - W_f(A)}.
\ee 
We first treat the Abelian gauge field as an external
classical
field and only consider the path integral over fermions. 
The second step is to take the resulting generating functional for the
gauge field and
combine it with the standard action for the gauge field and perform
the path integral over the gauge fields. The conceptually difficult
part is the computation of $W_f(A)$ and we will only concern ourselves
with this part. Our aim is obtain a correct lattice regularized
version for $W_f(A)$.
One can use standard numerical techniques to
perform the integral over the lattice gauge field.

The organization of the lectures is as follows. We will start by
describing the details of abelian gauge fields on a two dimensional
lattice
with periodic boundary conditions. This will help us understand the
Hodge decomposition of the gauge fields into torons, representative in
a gauge orbit, gauge transformations and a global topological piece.

We will take the continuum limit of the background gauge field and
couple a Dirac fermion to it. We will be able to solve the eigenvalue
problem using the Hodge decomposition. In particular, we will be
able to show the presence of chiral zero modes when the background 
gauge field has a global topological piece. Formally, the path
integral
over fermions is the product over the eigenvalues as long as there
are no zero modes and if one could properly regularize the Jacobian.
This will lead us to the problem of fermions on the lattice as a
possible regularization scheme.  

We will explain the presence of
lattice doublers and the solution of Wilson. But this will not help us
deal with chiral fermions on the lattice. In order to understand the
subtleties associated with regulating a chiral fermion, we will first
focus
on continuum Pauli-Villars regularization (we do not consider
dimensional
regularization since we are ultimately interested in the
non-perturbative
lattice regularization) and show the need for
an infinite number of Pauli-Villars fields to regulate a single chiral
fermion. We will only show that all diagrams are regulated but we will
show
that
the idea can be extended to the lattice and therefore outside
perturbation theory. This is called the overlap formalism. 

We will describe the overlap formalism in detail. We will derive the
formula for the generating function associated with a chiral fermion.
We will derive the massless overlap Dirac operator for the case of
vector like theories. We will show how the overlap formalism realizes
gauge field topology. 

Our discussion of chiral anomalies will start with the discussion of
consistent
and covariant current in the context of the overlap formalism. We will
show that the covariant current is unambiguously defined but the
consistent current is not. We will show that there is an unambiguous
curvature obtained from the difference of the two currents.
Vanishing curvature would imply an anomaly free chiral gauge theory
in the continuum. We will use torons to demonstrate cancellation of
anomalies in a chiral gauge theory. We will use specific gauge field
backgrounds to compute the covariant anomaly. We will also 
demonstrate gauge field topology on the lattice. All these will be
numerical computations to help us understand how to work with
overlap fermions and to also understand lattice artifacts in the
context
of overlap fermions.

\section{Abelian gauge fields in two dimensions}

Consider a two dimensional $L_1\times L_2$ lattice with
$(n_1,n_2)$; $0\le n_1< L_1$ and $0\le n_2 < L_2$ labeling
the sites on the lattice. Let $U_1(n_1,n_2)$ and $U_2(n_1,n_2)$
be the link variables on the links connecting $(n_1,n_2)$ with
$(n_1+1,n_2)$ and $(n_1,n_2+1)$ respectively. The link
variables are $U(1)$ values phases. 

Under a local gauge transformation,
\bea
U_1(n_1,n_2) &\to & U^\prime_1(n_1,n_2)= g^\dagger(n_1,n_2) U_1(n_1,n_2) g(n_1+1,n_2);\cr
U_2(n_1,n_2) &\to & U^\prime_2(n_1,n_2)= g^\dagger(n_1,n_2)
U_2(n_1,n_2) g(n_1,n_2+1),
\label{locgauge}
\eea
where
$g(n_1,n_2) \in U(1)$ satisfy periodic boundary conditions.
Define the plaquette variable as
\be
e^{iE(n_1,n_2)} = U_1(n_1,n_2) U_2(n_1+1,n_2) U_1^\dagger(n_1,n_2+1)
U_2^\dagger(n_1,n_2); \ \  -\pi < E(n_1,n_2) \le \pi;
\ee
and it is easy to show that it is invariant under a local gauge transformation.
Let 
\be
e^{i\pi h_1} = \prod_{n_1=0}^{L_1-1} U_1(n_1,0);\ \ \ 
e^{i\pi h_2} = \prod_{n_2=0}^{L_2-1} U_2(0,n_2);\ \ \  -1 < h_1,h_2 \le 1.
\ee
These two additional variables are called torons and
are also invariant under local gauge transformations.
The gauge action on the lattice is
\be
S_g = -\beta\sum_{n_1,n_2} \cos E(n_1,n_2).
\ee
In order to take the continuum limit on a torus of fixed physical
size, $l_1\times l_2$, we will take $\beta,L_1,L_2\to\infty$, keeping
$\frac{L_1}{\sqrt{\beta}} = l_1$ and $\frac{L_2}{\sqrt{\beta}}=l_2$ fixed.

\subsection{Hodge decomposition}

Since  a constant $g(n_1,n_2)$ does not change the link variables,
we set $g(0,0)=1$.
We can choose $g(n_1,0)$; $0< n_1 < L_1$
and $g(0,n_2)$; $0< n_2 < L_2$ such that 
\be
U^\prime_1(n_1,0)=e^{i\frac{\pi h_1}{L_1}};\ \ 0\le n_1 < L_1
\ee
\be
U^\prime_2(0,n_2)=e^{i\frac{\pi h_2}{L_2}};\ \ 0\le n_2 < L_2
\ee
Next, we choose the remaining $g(n_1,n_2)$ such that
\be
U^\prime_1(n_1,n_2) = e^{i\frac{\pi h_1}{L_1}};\ \ 0\le n_1 < L_1-1; \ \
0 < n_2 < L_2.
\ee
It is now clear that
\be 
U^\prime_2(n_1,n_2) = e^{i\sum_{k_1=0}^{n_1-1}E(k_1,n_2)}e^{i\frac{\pi h_2}{L_2}};\ \ \  0 < n_1 < L_1;\ \ 0\le n_2
< L_2,
\ee 
and
\be
U^\prime_1(L_1-1,n_2) =
e^{-i\sum_{k_2=0}^{n_2-1}\sum_{k_1=0}^{L_1-1}E(k_1,k_2)}
e^{i\frac{\pi h_1}{L_1}};
\ \  
 0<n_2< L_2.
\ee
Since the above procedure did not require
$E(L_1-1,L_2-1)$, we conclude that all plaquettes variables are not
independent variables and
they satisfies the condition
\be
\sum_{k_1=0}^{L_1-1}\sum_{k_2=0}^{L_2-1} E(k_1,k_2) = 2\pi Q  \label{qtop}
\ee
where $Q$ is an integer, referred to as the topological charge

Let us now assume we are given the gauge invariant degrees of freedom, namely,
\begin{itemize}
\item $(L_1L_2)$ electric flux degrees of freedom $E(n_1,n_2)$ that satisfy
  (\ref{qtop});
\item and the two toron variables, $h_1$ and $h_2$.
\end{itemize}
The full set of gauge fields consistent with the above data are
\bea
U_1(n_1,n_2) &=& e^{i\frac{\pi h_1}{L_1}} U^Q_1(n_1,n_2)
e^{-i\chi(n_1,n_2) }e^{i\left[\phi(n_1,n_2-1)-\phi(n_1,n_2)\right]}
  e^{i\chi(n_1+1,n_2)}\cr
U_2(n_1,n_2) &=& e^{i\frac{\pi h_2}{L_2}} U^Q_2(n_1,n_2)
e^{-i\chi(n_1,n_2) }e^{i\left[\phi(n_1,n_2)-\phi(n_1-1,n_2)\right]}
  e^{i\chi(n_1,n_2+1)}
\label{u1u2lat}
\eea
where $\phi(n_1,n_2)$ is the solution to
\bea
\Box\phi(n_1,n_2) &=& 
-4\phi(n_1,n_2)+\phi(n_1-1,n_2)+\phi(n_1+1,n2)+\phi(n_1,n_2-1)+\phi(n_1,n_2+1)\cr
&=& E(n_1,n_2) - \frac{2\pi Q}{L_1L_2},\label{lphieqn}
\eea
and 
\bea
U_1^Q(n_1,n_2) &=& e^{-i\frac{2\pi Q}{L_1 L_2} n_2};\ \ \  0\le n_1 <
L_1;\ \ 0\le n_2 <
L_2;\cr
U_2^Q(n_1,n_2) &=& \cases{ 1 & if $0 \le n_1 < L_1$ and $0 \le n_2 <  L_2-1$\cr
 e^{i\frac{2\pi Q}{L_1} n_1} & if $0\le n_1 < L_1$ and $n_2=L_2$}\label{utop}
\eea
is the topological part of the gauge field.

In the continuum limit, 
\be
U_\mu(n_1,n_2) \to e^{\frac{i}{\sqrt{\beta}}A_\mu(x_1,x_2)};\ \ \ \
x_\mu = \frac{n_\mu}{\sqrt{\beta}}.
\ee
In the continuum limit, (\ref{utop}), leads to a discontinuous function for
$A_\mu(x_1,x_2)$
and it is natural to move the discontinuity to a gauge transformation
that is not periodic on the torus. 
The gauge field in the continuum limit is written as
\bea
A_1(x_1,x_2) &=& \frac{\pi h_1}{l_1} + \partial_1 \chi(x_1,x_2) -\partial_2\phi(x_1,x_2) -
\frac{2\pi Q}{l_1l_2}x_2\cr 
A_2(x_1,x_2) &=& \frac{\pi h_2}{l_2} + \partial_2 \chi(x_1,x_2)
+\partial_1\phi(x_1,x_2) .
\label{hodgec}
\eea
Clearly, $A_1(x_1,0)\ne A_1(x_1,l_2)$, but
\be
A_1(x_1,l_2) = A_1(x_1,0) - \partial_1 \frac{2\pi Q x_1}{l_1},\label{winding}
\ee
referred to as a gauge transformation on the torus with non-trivial
winding.
The continuum limit of (\ref{lphieqn}) is
\be
\Box\phi(x_1,x_2)= \left( \partial_1^2 + \partial_2^2\right)\phi(x_1,x_2) = e(x_1,x_2) -\frac{2\pi
  Q}{l_1l_2},\label{philap}
\ee
where $\beta E(n_1,n_2) \to e(x_1,x_2)$ is the continuum limit of the
electric field.
\subsection {Fermion zero modes}\label{fzm}
The massless Dirac operator coupled to the abelian gauge field is
given by
\be
\Feyn{D} = \sum_{\mu=1}^2 \sigma_\mu (\partial_\mu + i A_\mu).
\ee
We will use the chiral representation for the Pauli matrices, namely,
\be
\sigma_1=\pmatrix{0 & 1\cr 1 & 0\cr};\ \ \ 
\sigma_2=\pmatrix{0 & -i\cr i & 0\cr};\ \ \ 
\sigma_3 = -i\sigma_1\sigma_2 = \pmatrix{ 1 & 0 \cr 0 & -1 \cr}.
\ee
The chiral symmetry is due to the identity,
\be
\Feyn{D} \sigma_3 = -\sigma_3 \Feyn{D},
\ee
since eigenvalues of $\Feyn{D}$ comes in $\pm i\lambda$ pairs.
In addition, the Dirac operator always has
exactly $|Q|$ zero modes with definite chirality (eigenvectors of
$\sigma_3$).
These zero modes are robust and are present for all gauge
fields with a fixed $Q$ since
\be
\Feyn{D} =f^*e^{-\sigma_3\phi}\Feyn{D}_Q
fe^{-\sigma_3\phi},
\ee
where
\be
f=e^{i\frac{\pi h_1x_1}{l_1}+i\frac{\pi h_2x_2}{l_2} +i\chi},
\ee
and
\be
\Feyn{D}_Q=\pmatrix { 0 & C_Q \cr
-C^\dagger_Q & 0
 \cr};\ \ \ 
C_Q=\partial_1 - i \partial_2 -ibx_2;\ \ \ b=\frac{2\pi Q}{l_1 l_2} ,
\ee

To solve for the zero modes of $\Feyn{D}_Q$, note that
\be
-\Feyn{D}^2_Q = \pmatrix { C_Q C^\dagger_Q & 0 \cr 0 & C^\dagger_Q
  C_Q\cr} =
\pmatrix {  K - b & 0 \cr
0 &  K+ b\cr},
\ee
where
\be
K=-\partial_1^2 -\partial_2^2 +2i b x_2 \partial_1 + b^2
  x_2^2 .
\ee
In order to make the Dirac equation gauge covariant in the presence of
non-trivial winding as given in (\ref{winding}), 
we are interested in solutions to
\be
K \pm b \psi =0,
\ee
with the fermion
wave-functions satisfying the boundary conditions
\be
\psi(x_1+l_1,x_2)=\psi(x_1,x_2);\ \ \  \psi(x_1,x_2+l_2) =
e^{i\frac{2\pi Q}{l_1}x_1}\psi(x_1,x_2),\label{zerobound}
\ee
on the torus.
Solutions have to be of form
\be
\psi(x_1,x_2) = \sum_{p=-\infty}^\infty a_p e^{i\frac{2\pi p}{l_1} x_1} 
h\left(x_2-\frac{pl_2}{Q}\right)
\ee 
where $h(y)$ is the solution to the Harmonic Oscillator,
\be
\left( -\partial_y^2 +b^2y^2 \pm b\right) h(y) =0.
\ee
The normalizable solution is
\be
h(y) = e^{-\frac{|b|}{2} y^2},
\ee
and has positive chirality if $Q>0$ and negative chirality if $Q<0$.
Our solutions are therefore of the form,
\be
\psi(x_1,x_2) = \sum_{p=-\infty}^\infty a_p e^{i\frac{2\pi p}{l_1} x_1} 
e^{-\frac{\pi|Q|}{l_1l_2}\left(x_2-\frac{pl_2}{Q}\right)^2}.
\ee 
Since
\be
\psi(x_1,x_2+l_2) = \sum_{p=-\infty}^\infty a_{p+Q} e^{i\frac{2\pi (p+Q)}{l_1} x_1}
e^{-\frac{\pi|Q|}{l_1l_2}\left(x_2-\frac{pl_2}{Q}\right)^2},
\ee
we see that 
\be
a_{p+Q} = a_p,
\ee
for all $p$ in order for the solution to satisfy the boundary
conditions in (\ref{zerobound}).
Therefore, we have exactly $|Q|$ zero modes with definite chirality.
The orthogonal set of solutions are
\bea
\psi_k(z_1,z_2) &=& \frac{1}{\sqrt{l_1l_2} }\sum_{p=-\infty}^\infty e^{i2\pi (pQ+k)z_1}
e^{-\pi|Q|\tau\left(z_2-p-\frac{k}{Q}\right)^2}\cr
&=& \frac{1}{\sqrt{l_1l_2}} e^{2\pi i k z_1 -\pi |Q|\tau
    \left(z_2-\frac{k}{Q}\right)^2}
\vartheta\left(Q\left(z_1+i\tau z_2\right) -ik\tau; i|Q|\tau\right),\label{zeromodes}
\eea
for $k=0,\cdots,|Q-1|$ and
$z_\mu = \frac{x_\mu}{l_\mu}$, $\tau=\frac{l_2}{l_1}$.

The zero modes define an index since 
$C_Q^\dagger$ has $Q$ zero modes and $C_Q$ has none for $Q>0$
and vice-versa. It is not possible to realize this using a finite
matrix
for $C_Q$. The need for an infinite number of degrees of freedom to
properly realize a single chiral fermions becomes clear and we will
see this to be the case even at the level of continuum perturbation theory.

\section{Wilson Fermions}

Before, we proceed with the continuum regularization of chiral
fermions,
we briefly digress to understand a well known problem with
realizing fermionic degrees of freedom on the lattice.

Since $E(n_1,n_2)$ and $\phi(n_1,n_2)$ are periodic functions, we can
write them in terms of their Fourier components:
\be 
E(n_1,n_2) -\frac{2\pi Q}{L_1L_2} = \frac{1}{L_1L_2}
\sum_{p_1=-\flfn{\frac{L_1-1}{2}}}^{\clfn{\frac{L_1-1}{2}}}
\sum_{p_2=-\flfn{\frac{L_2-1}{2}}}^{\clfn{\frac{L_2-1}{2}}}
\tilde E(p_1,p_2) 
e^{i\frac{2\pi p_1n_1}{L_1} + i\frac{2\pi p_2n_2}{L_2}},
\ee 
and
\be
\phi(n_1,n_2)  = \frac{1}{L_1L_2}
\sum_{p_1=-\flfn{\frac{L_1-1}{2}}}^{\clfn{\frac{L_1-1}{2}}}
\sum_{p_2=-\flfn{\frac{L_2-1}{2}}}^{\clfn{\frac{L_2-1}{2}}}
\tilde \phi(p_1,p_2)
e^{i\frac{2\pi p_1n_1}{L_1} + i\frac{2\pi p_2n_2}{L_2}}.
\ee
(\ref{qtop}) implies that $\tilde E(0,0)=0$ and
(\ref{lphieqn}) becomes
\be
\tilde
\phi(p_1,p_2) = \cases{
0 & if $p_1=p_2=0$ \cr
-\frac{\tilde E(p_1,p_2)}{4\left[\sin^2(\frac{\pi p_1}{L_1})
    +\sin^2(\frac{\pi p_2}{L_2})\right]}
& otherwise}.\label{phisol}
\ee

In the continuum limit, $p_\mu$, take on all integer values and since
$\beta\tilde E(p_1,p_2) \to \tilde e(p_1,p_2)$, we see that
(\ref{phisol})
correctly goes to the continuum limit of the solutions to
(\ref{philap})
for all finite values of $p_\mu$. The ultraviolet behavior is modified
by the lattice regulator but the denominator in (\ref{phisol}) which
is
the propagator of a Klein-Gordon field
remains a monotonic function of the momenta.
In particular, the only pole in the propagator in a Brillouin zone,
$p_\mu \in \left[-\flfn{\frac{L_\mu-1}{2}}, \clfn{\frac{L_\mu-1}{2}}
\right]$
is at $p_1=p_2=0$.

Let $\psi(n_1,n_2)$ be a Dirac field in two dimensions. In analogy
with (\ref{lphieqn}), we can write the Dirac operator as
\be
\Feyn{D} \psi(n_1,n_2)= \sigma_1 [\psi(n_1+1,n_2)-\psi(n_1-1,n_2)]+
\sigma_2 [\psi(n_1,n_2+1)-\psi(n_1,n_2-1)],
\ee
which is
\be
\Feyn{D} \tilde\psi(p_1,p_2) = 2i\left[ \sigma_1\sin \left(\frac{2\pi p_1}{L_1}\right)
+\sigma_2\sin \left(\frac{2\pi p_2}{L_2}\right) \right]\tilde\psi(p_1,p_2) 
\ee
after a Fourier transform.
Contrary to the case of Klein-Gordon field, the propagator of a Dirac
field is not a monotonic function of the momenta and has additional
zeros at the boundaries of the Brillouin zero when $L_\mu\to\infty$.
Therefore, we will end up with four massless Dirac fermions in the
continuum limit of a two dimensional theory. 
This is referred to as the ``doubling phenomenon''. 

The number of particles do not change if we add a mass term but we
could give masses to the particles at the boundary of the
Brillouin zone keeping the fermion massless at the center
of the Brillouin zone. If the masses for the particles
at the boundary become infinite in the continuum limit, we will only
have one massless fermion in the continuum limit and we could
treat the infinitely massive fermions as regulator fields.
In fact the simplest choice for a momentum dependent mass term
is $-\Box\psi$. The massless Wilson-Dirac operator is
\be
\Feyn{D}_w = \Feyn{D} - \Box,\label{wilson}
\ee
and will describe a single massless Dirac fermion in the continuum
limit as can be seen by taking the limit of $\sqrt\beta \Feyn{D}_w$
in momentum space for momenta inside the Brillouin zone
 and for momenta in the boundary of the Brillouin zone.

Whereas the Wilson-Dirac operator solves the problems of doublers,
adding a mass term breaks the chiral symmetry, namely,
\be 
\sigma_3 \Feyn{D} \sigma_3 = -\Feyn{D}.
\ee
In particular, it does not solve the problem when we want to write
down a theory with Weyl
fermions,
namely,
\be
\Feyn{C} =  \Feyn{D} P_+ = P_- \Feyn{D},
\ee
where
\be
P_\pm = \frac{1\pm \sigma_3}{2}.
\ee
 In order to understand the solution to the problem, we will need to
review chiral anomalies and fermion zero modes in a gauge field
background with non-zero topological charge.

\section{Continuum Pauli-Villars regularization}\label{cpvr}

Let $\psi_\pm(x)=P_\pm\psi(x)$ and $\ub\psi_\pm(x) = \ub\psi(x) P_\mp$
denote a left(+) or right(-)
handed fermion and anti-fermion at the Euclidean point $x=(x_1,x_2)$.
It will be useful to define $\partial = \partial_1 + i \partial_2$ and
$\ub\partial = \partial_1 - i \partial_2$. It will also be useful to
define, $A=A_1+iA_2$ and $\ub A = A_1 - i A_2$.
The fermion action for a left handed field of charge $q_+$ is
\be
S_{+f}(\ub\psi_+,\psi_+,A) 
= \int d^2x \ub\psi_+ \partial
\psi_+ + i q_+ \int d^2 x \ub\psi_+A\psi_+\label{leftaction}
\ee
and the fermion action for a right handed field of charge $q_-$ is
\be
S_{-f}(\ub\psi_-,\psi_-,A) = \int d^2x \ub\psi_- \ub\partial
\psi_- + i q_- \int d^2 x \ub\psi_- \ub A\psi_-\label{rightaction}
\ee

Using the Hodge decompostion in (\ref{hodgec}) and setting $ h_\mu=0$
and $Q=0$
for the moment, we can write
\be
 A = \partial \xi;\ \ \ \ \xi=\chi+i\phi.\label{atop0}
\ee

Our aim is to compute the path integral over the fermion fields for
a fixed background $\xi$.
Consider the result of the path integral over left handed fermions:
\be
e^{-W_+(A)} = \int [d\ub\psi][d\psi] e^{S_{+f}(\ub\psi_+,\psi_+,A) }.\label{unreglef}
\ee
The change of variables,
\be
\psi_+ \to \psi^\prime_+ = e^{i\xi} \psi_+;\ \ \ 
\ub\psi_+ \to \ub\psi^\prime_+ = e^{-i\xi} \ub\psi_+;\ \ \ 
\psi_- \to \psi^\prime_- = e^{i\ub\xi} \psi_-;\ \ \ 
\ub\psi_- \to \ub\psi^\prime_- = e^{-i\ub\xi} \ub\psi_-,
\ee
formally decouples the fermion action from the gauge field.
Any dependence on $\xi$ due to the integration over the fermion fields
should arise from the Jacobian.
The Jacobian associated with such a transformation is not unity
since there are divergences associated
with the path integral.

To see the above argument in standard perturbation theory of (\ref{unreglef}),
we write
\bea
\psi(x) &=& \frac{1}{\sqrt{l_1l_2}}\sum_{p_\mu=-\infty}^\infty \tilde\psi(p) e^{i\frac{2\pi
  p_1x_1}{l_1}+i\frac{2\pi p_2x_2}{l_2}}\cr
\ub\psi(x) &=& \frac{1}{\sqrt{l_1l_2}}\sum_{p_\mu=-\infty}^\infty \ub{\tilde\psi}(p) e^{-i\frac{2\pi
  p_1x_1}{l_1}-i\frac{2\pi p_2x_2}{l_2}},\cr
A(x) &=& \sum_{p_\mu=-\infty}^\infty \tilde A(p) e^{i\frac{2\pi
  p_1x_1}{l_1}+i\frac{2\pi p_2x_2}{l_2}}.\label{fourtran}
\eea
Note that (\ref{atop0}) implies that $\tilde A_\mu(0)=0$.

The free fermion propagator is
\be
\langle \tilde\psi_+(q)\ub{\tilde\psi}_+(p)\rangle
=-\frac{\delta_{pq}}{{\cal P}},\ \ \  {\cal P} = 2\pi i \left( \frac{p_1}{l_1}+i\frac{p_2}{l_2}\right),
\ee
and the interaction vertex is 
\be
S_I=iq_+\sum_{p_\mu,q_\mu=-\infty}^\infty \ub{\tilde\psi}_+(p+q) \sigma_\mu
\tilde A_\mu(p) \tilde\psi_+(q).
\ee
The contributions to $W_+(A_\mu)$ correspond to a single fermion loop
with several insertions of $A_\mu$:
\bea
W_+(A_\mu) = \sum_{k=1}^\infty (-iq_+)^k && \sum_{p_\mu^j=-\infty}^\infty
\left[ \prod_{j=1}^k \tilde
A(p^j) \right] \cr
&& \sum_{r_\mu=-\infty}^\infty
\frac{1}{{\cal R}}
\frac{1}{{\cal R}+{\cal P}^1}
\frac{1}{{\cal R}+{\cal P}^1+{\cal P}^2}\cdots
\frac{1}{{\cal R}+{\cal P}^1+{\cal P}^2+\cdots+{\cal P}^{k-1}}\cr
&&
\delta\left(\sum_{j=1}^k p^j\right).\label{wlamu}
\eea
There is no contribution from $k=1$ since $\tilde A(0)=0$.
For $k>2$, the sum over $r_\mu$ will be convergent and we can
rearrange the sum as follows. We can write each term as a sum
over simple poles and the sum of the residues will be zero.
We can shift the sum over $r_\mu$, such that all poles are at zero.
Since the sum over residues is zero, the sum will be zero.
There is no contribution to $W_+(A_\mu)$ from $k>2$.

In order to compute the contribution from $k=2$,
we will use a Pauli-Villars regularization that acts
independently on the left and right handed fields and fully regulates
the combined contribution from the left and right handed fields
provided the full set of fields satisfy the anomaly cancellation
condition:
\be
\sum_{i=1}^{n_+} \left(q_+^i\right)^2 = 
\sum_{j=1}^{n_-} \left(q_-^j\right)^2.\label{anomcan}
\ee 
If $n_+=n_-$ and $q_+^i=q_-^i$ for all $i$, we have a QCD-like
theory and this can be regulated in the standard manner using
a finite number of Pauli-Villars fields. If the left handed field
content
is not the same as the right handed field content, we need to
have Pauli-Villars fields associated with a single left or right
handed
field. Since Pauli-Villars fields have a mass and have twice the
number
of degrees of freedom compared to a single left or right handed field,
a finite number of Pauli-Villars fields is not sufficient to regulate
a single component. We therefore assume that the regulated path integral
associated with (\ref{unreglef}) is of the form
\be
e^{-W_+(A_\mu)} = \int [d\ub\psi][d\psi]
e^{\int d^2x  \ub\psi\left[  \sigma_\mu (\partial_\mu +
  iq_+A_\mu)
+P_+ M 
+P_- M^\dagger \right]\psi},\label{leftreg}
\ee
where $\psi$ represents an infinite set of Dirac fields, $\psi_n$,
$n=0,\cdots,\infty$.
Every Dirac field couples to the same gauge field and carries the same
charge.

Under parity, $x_1\to -x_1$ and $x_2\to x_2$,
\be
A_1(x_1,x_2) \to -A_1(-x_1,x_2);\ \   A_2(x_1,x_2) \to A_2(-x_1,x_2),
\ee
and
\be
\psi(x_1,x_2) \to \sigma_2\psi(-x_1,x_2);\ \ 
\ub\psi(x_1,x_2) \to \ub\psi(-x_1,x_2) \sigma_2;
\ee
and we see that the action for a left handed field in (\ref{leftaction})
goes into the action for a right handed field in (\ref{rightaction})
with the same charge. 
We also see that the regularized action in (\ref{leftreg})
goes into an action with 
$M\to M^\dagger$. 
We will requite
$M$ to be an infinite dimensional complex mass matrix  
such that $M$ has one zero mode and $M^\dagger$ has no zero mode 
in order to correctly describe  a single left handed fermion. 
Under parity, the zero mode shifts from one
chirality to other as expected.

\subsection{Free fermion propagator:}
The free fermion part of the action in (\ref{leftreg}) is
\be
S_q = 
\int d^2x  \pmatrix{\ub\psi_+(x) & \ub\psi_-(x)\cr}
\pmatrix{ \partial & 
M^\dagger \cr 
M & \ub\partial\cr } 
\pmatrix{\psi_+(x)\cr\psi_-(x)\cr}
\ee
which is
\be 
S_q = \sum_{p_\mu=-\infty}^\infty 
\pmatrix{ \ub{\tilde\psi}_+(p) & \ub{\tilde\psi}_-(p)\cr} 
\pmatrix{ {\cal P}_1+i{\cal P}_2 & 
M^\dagger \cr 
M & {\cal P}_1-i{\cal P}_2\cr } 
\pmatrix{\tilde\psi_+(p) & \tilde\psi_+(p)\cr};
\ \ \ {\cal P}_\mu = 2\pi i \frac{p_\mu}{l_\mu}
\ee 
using (\ref{fourtran}). 
The interaction vertex in momentum space is
\be 
S_I=iq_+\sum_{p_\mu,q_\mu=-\infty}^\infty 
\left[ \ub{\tilde\psi}_+(p+q) \tilde A(p) \tilde\psi_+(q)+
\ub{\tilde\psi}_-(p+q) \ub{\tilde A}(p) \tilde\psi_-(q)\right],
\ee 
using (\ref{fourtran}).

The free fermion propagator is
\bea
\langle\tilde\psi_+(p)\ub{\tilde\psi}_+(p)\rangle = \frac{{\cal
   P}_1-i{\cal P}_2}{|{\cal P}|^2 +M^\dagger M}, &&
\langle\tilde\psi_-(p)\ub{\tilde\psi}_-(p)\rangle = \frac{{\cal
   P}_1+i{\cal P}_2}{|{\cal P}|^2 +MM^\dagger },\cr
\langle\tilde\psi_-(p)\ub{\tilde\psi}_+(p)\rangle = -M\frac{1}{|{\cal P}|^2 +M^\dagger M},&&
\langle\tilde\psi_+(p)\ub{\tilde\psi}_-(p)\rangle = -M^\dagger \frac{1}{|{\cal P}|^2 +MM^\dagger},
\eea
where $|{\cal P}|^2 = -{\cal P}_1^2 -{\cal P}_2^2$.

As a specific example, let us take the infinite dimensional space to
be labelled by a discrete index, $i=0,1,\cdots\infty$. Let 
\be 
M_{jk} = \Lambda k \delta_{j,k-1}; \ \ \
M^\dagger_{jk} = \Lambda j \delta_{j,k+1}; 
\ee 
where $\Lambda$ is the regulator mass to be taken to $\infty$. 
The zero mode of $M$ is 
\be
v_k=\cases{1 & if $k=0$ \cr 0 & otherwise};\ \ \  Mv=0,
\ee
and there is no zero mode for $M^\dagger$.
Furthermore,
\be
(MM^\dagger)_{jk} = \Lambda^2(j+1)^2 \delta_{jk}; \ \ \ 
(M^\dagger M)_{jk} = \Lambda^2 j^2 \delta_{jk}.
\ee
Explicit expressions for the propagators are
\bea
G^{++}_{jj}(p) = \frac{{\cal P}_1-i{\cal P}_2}{{\cal P}^2
  +\Lambda^2 j^2} &&
G^{--}_{jj}(p)=\frac{{\cal P}_1+i{\cal P}_2}{{\cal P}^2+\Lambda^2(j+1)^2};\cr
G^{-+}_{j(j+1)}(p) = -\frac{\Lambda(j+1)}{{\cal
    P}^2+\Lambda^2(j+1)^2}; &&
G^{+-}_{(j+1)j}(p) =  -\frac{\Lambda(j+1)}{{\cal P}^2+\Lambda^2(j+1)^2};
\eea
for $j=0,\cdots,\infty$.
The only massless particle is one with positive chirality and all
other poles in the propagator correspond to the regulator poles.

\subsection{Computation of $W_+(A_\mu)$:}

The spin-statistics is assigned as follows to successfully regulate
the theory:
\bea
\ub\psi_{+j};\ \ \psi_{+j}:&&\cases{ & fermion for even $j$\cr & boson
for odd $j$};\cr
\ub\psi_{-j};\ \ \psi_{-j}:&&\cases{ & fermion for odd $j$\cr & boson
for even $j$}.\label{spinstat}
\eea
Only the $k=2$ term contributes to (\ref{wlamu}) which can be written
as
\bea
W_+(A_\mu) &=& q_+^2\sum_{p_\mu=-\infty}^\infty\cr
&& \tilde A(p) {\tilde A}(-p) \sum_{q_\mu=0}^\infty 
 \sum_{j=0}^\infty (-1)^j G^{++}_{jj}(q-\frac{p}{2})G^{++}_{jj}(q+\frac{p}{2})\cr
&&+\ub{\tilde A}(p) \ub{\tilde A}(-p) \sum_{q_\mu=0}^\infty
\sum_{j=0}^\infty (-1)^{j+1} G^{--}_{jj}(q-\frac{p}{2})G^{--}_{jj}(q+\frac{p}{2})\cr
&&+{\tilde A}(p) \ub{\tilde A}(-p) \sum_{q_\mu=0}^\infty 
\sum_{j=0}^\infty (-1)^j G^{+-}_{j(j+1)}(q-\frac{p}{2})G^{-+}_{(j+1)j}(q+\frac{p}{2})\cr
&&+\ub{\tilde A}(p) {\tilde A}(-p) \sum_{q_\mu=0}^\infty 
\sum_{j=0}^\infty (-1)^j G^{-+}_{(j+1)j}(q-\frac{p}{2})G^{+-}_{j(j+1)}(q+\frac{p}{2})\label{vacpol}
\eea
The sum over the internal momenta, $q_\mu$, convergent term by term in
$j$ for the expressions in the last two lines of (\ref{vacpol}). We
focus on the first two terms and write the intermediate result before
we
perform the sum over $p_\mu$ and $q_\mu$ as
\bea
&q_+^2
\Biggl\{\left[ \tilde A_1(p) \tilde A_1(-p) - \tilde A_2(p) \tilde A_2(-p) \right]
\left[ 
({\cal Q}_1-\frac{{\cal P}_1}{2}) ({\cal Q}_1+\frac{{\cal P}_1}{2})
-({\cal Q}_2-\frac{{\cal P}_2}{2}) ({\cal Q}_2+\frac{{\cal
P}_2}{2})\right]&\cr
&+\left[ \tilde A_1(p) \tilde A_2(-p) + \tilde A_2(p) \tilde A_1(-p) \right]
\left[ 
({\cal Q}_1-\frac{{\cal P}_1}{2}) ({\cal Q}_2+\frac{{\cal P}_2}{2})
+({\cal Q}_2-\frac{{\cal P}_2}{2}) ({\cal Q}_1+\frac{{\cal
P}_1}{2})\right]\Biggr\}&\cr
&\sum_{j=-\infty}^\infty
\frac{(-1)^j}{\left[\left({\cal Q}-\frac{\cal
 P}{2}\right)^2 + \Lambda^2 j^2\right]
\left[\left({\cal Q}+\frac{\cal
 P}{2}\right)^2 + \Lambda^2 j^2\right]}&\cr
+&iq_+^2 
\Biggl\{\left[ \tilde A_2(p) \tilde A_2(-p) - \tilde A_1(p) \tilde A_1(-p) \right]
\left[ 
({\cal Q}_1-\frac{{\cal P}_1}{2}) ({\cal Q}_2+\frac{{\cal P}_2}{2})
+({\cal Q}_2-\frac{{\cal P}_2}{2}) ({\cal Q}_1+\frac{{\cal
P}_1}{2})\right]&\cr
&+\left[ \tilde A_1(p) \tilde A_2(-p) + \tilde A_2(p) \tilde A_1(-p) \right]
\left[ 
({\cal Q}_1-\frac{{\cal P}_1}{2}) ({\cal Q}_1+\frac{{\cal P}_1}{2})
-({\cal Q}_2-\frac{{\cal P}_2}{2}) ({\cal Q}_2+\frac{{\cal
P}_2}{2})\right]\Biggr\}&\cr
&\frac{1}{\left[\left({\cal Q}-\frac{\cal
 P}{2}\right)^2 \right]
\left[\left({\cal Q}+\frac{\cal
 P}{2}\right)^2 \right]}\label{pvreg}
\eea

If we repeat the calculation for a regulated right handed fermion, we
would have interchanged $M$ and $M^\dagger$ in (\ref{leftreg}) and
we would have also interchanged the statistics in (\ref{spinstat}).
Working through the intermediate steps, we would conclude that
the we would have obtained the complex conjugate of the
expression in (\ref{pvreg}) along with $q_+\to q_-$.
If we satisfy, (\ref{anomcan}), then the resulting expression would
be real. Since
\be
\sum_{j=-\infty}^\infty \frac{(-1)^j}{{\cal R}^2 + \Lambda^2 j^2}
= \frac{\pi}{\Lambda |{\cal R}| \sinh \left(\frac{\pi |{\cal
        R}|}{\Lambda}\right)},\label{convtow}
\ee
goes to zero exponentially fast as $|{\cal R}| \to \infty$, the
sum over $q_\mu$ will converge and result in a finite expression for
$W(A_\mu)$ for the anomaly free chiral gauge theory.
Since each chiral fermion has its own infinite tower of Pauli-Villars
regulator fields, the chiral symmetry that acts on each chiral
fermion is still a symmetry of the regulated theory.

\section{Overlap formalism}

Our aim is to extend this idea to define a non-perturbative
regularization
of anomaly free chiral gauge theories. 
It is best to think of the infinite set of fields as an extra
dimension and view the fermion fields as $\psi(x,s)$
and $\ub\psi(x,s)$. Our example in the previous section corresponds to
\be
M=\frac{\Lambda}{2}(\partial_s + s)\sqrt{-\partial_s^2+s^2-1}.
\ee
Let us view $s$ as time and consider the many body Hamiltonian associated with
(\ref{leftreg}).
The Hamiltonian will be quadratic in the fermion (boson) creation and
annihilation
operator and will conserve particle number. The path integral is
regulated
because the ground state energy was {\sl arranged to be zero} for our
choice of the set of fermions and bosons. The ground state as
$s\to\infty$
is different from the ground state as $s\to-\infty$ due to the $s$
dependent
operator $M$. The result of the path integral is simply an overlap of
these two ground states.

We make the above idea explicit by considering the following set up.
We choose
\be 
M=\partial_s +m(s);\ \ \  m(s) = \cases{\Lambda & for $s > 0 $\cr -m  & for
 $s\le 0$\cr};\ \ \  m,\Lambda > 0.\label{domainwall}
\ee 
Clearly, $M$ has a zero mode in the space of normalizable funtions,
namely,
\be 
\phi(s) = \cases { e^{-\Lambda s} & for $s \ge 0$\cr e^{m s} & for $
  s\le 0$\cr}.
\ee
but $M^\dagger$ does not have a zero mode. 
In order to deal with spin-statistics, we first set
 $s\in [-l_s,l_s]$ with the aim of taking it to infinite 
at the end.  Let the fermions with the above $M$ operator be referred
to as $a$-type.
We consider two additional sets of fermions with $s\in
[-\frac{l_s}{2},\frac{l_s}{2}]$
and
\be  
M=\partial_s + m(s); \ \ \   m(s)=-m,
\ee
 for one set ($b$-type) and
\be 
M=\partial_s +m(s); \ \ \  m(s)=\Lambda,
\ee
for the other set ($c$-type).

The action in (\ref{leftreg}) for a given $M$ can be written as
\be 
S_f = \int d^2x ds \ub\psi \left [ \sum_{\mu=1}^3 \sigma_\mu (\partial_\mu + i 
q_+ A_\mu(x)) + m(s) \right] \psi.
\ee 
where the fields, $\ub\psi$ and $\psi$ are one of three types of fermions
depending upon the choice of $m(s)$.
In all three cases,
we have a three dimensional operator with a  specific three 
dimensional gauge field 
background that is obtained from the two dimensional gauge field background
($A_1(x)$; $A_2(x)$; $A_3=0$) and with 
mass term, namely, $m(s)$.
It is best to use the second quantized picture to work out the result
of the path integral since there is no gauge field in the third
direction and the gauge fields in the two directions do not depend on
the
third direction.
The two many body Hamiltonians associated with the path integrals  
with $s$ playing the role of time are
\be
{\cal H}^\pm = a^\dagger H^\pm a;\ \ \  
H^\pm = \sigma_3 \left[ \sum_{\mu=1}^2 \sigma_\mu\left(\partial_\mu +i
  A_\mu\right) +m^\pm \right] ; \ \  
m^+=\Lambda; \ \ \  m^-=-m.
\ee 
Let $\Lambda_\pm$ and $|\pm\rangle$ 
be the highest eigenvalues and the corresponding eigenvectors of
${\cal H}_\pm$.
 The $a$ type see both Hamiltonians, $b$ sees only ${\cal
 H}^-$ and $c$ sees only ${\cal H}^+$.  
Our aim is to take $l_s$ to infinity with free boundary conditions in the
two ends but with the condition that 
\hfill\break
$a$ type sees
$|+\rangle$ as the boundary state on the positive $s$ side
and
$|-\rangle$ as the boundary state on the negative $s$ side;
\hfill\break
$b$ type  sees
$|-\rangle$ as the boundary state on both ends;
\hfill\break
$c$ type  sees
$|+\rangle$ as the boundary state on both ends.
\hfill\break
Let $e^{W_{a,b,c}(A_\mu)}$ be the result of the path integral over the
$a,b,c$ type fermions respectively. Then,
\be
e^{W_a(A_\mu)} = e^{l_s\left(\Lambda_++\Lambda_-\right)}
\langle-|+\rangle;\ \ \ \ 
e^{W_b(A_\mu)} = e^{l_s \Lambda_-};
\ \ \ \ 
e^{W_c(A_\mu)} = e^{l_s \Lambda_+}.
\ee
We set the regulated determinant of a single left handed Weyl fermion
to
\be
e^{W_+(A_\mu)}= \lim_{l_s\to\infty} \frac{e^{W_a(A_\mu)}}{e^{W_b(A_\mu)} e^{W_b(A_\mu)}} =
    \langle - | +\rangle\label{overlapl}
\ee
and this is the overlap formula. Since we divided by the result of the
integral for$b$ and $c$ fermions, their statistics is opposite to that
of the $a$ type fermions. Their purpose is the same as the
Pauli-Villars regulator fields in the continuum formalism.
They subtract the contributions coming from the non-zero modes of
$M^\dagger M$.
It turns out we can make one more
simplification and take $\Lambda\to\infty$ since it does not affect
our zero mode condition nor does it affect the dependence of the gauge
field in an physical way as we will see.

The overlap formula trivially extends to any even dimensional chiral
gauge theory.
We will show in what follows that the overlap formula has all the
correct continuum properties desired of a well regulated chiral
determinant. 

A lattice realization of the overlap formula is simple since $H^-$ is
the
Hermitian massive Dirac operator and we can replace it by
\be
H_w=\sigma_3 (\Feyn{D}_w - m)
\ee
 using (\ref{wilson}).
We will assume this to be the case from now on and assume that
$H^+=\sigma_3$ ($\Lambda\to\infty$) and $H^-=H_w$.
We will also assume that the background gauge field are given
by the lattice link variables, $U_\mu(x)$.

\subsection{Phase of the overlap}\label{chiralinf}

The highest states of ${\cal H}^\pm$ are obtained by filling the
positive eigenstates of $H^\pm$ respectively. For $U_\mu(x)=1$, $H_w$
will have equal number of positive and negative eigenvalues and
this will remain true in perturbation theory. 
For definiteness, let us set the size of the Hamiltonians to be $2n$.
and set $\psi^\pm_k$, $k=1,\cdots, n$, as the positive
eigenstates of $H^\pm$ respectively.  Then,
\be
  \langle  - | +\rangle
= \det O;\ \ \  O_{jk} = [\psi^+_j]^\dagger \psi^-_k.
\ee

Since $H_+=\sigma_3$, it is diagonal with
$$
\pmatrix{1 \cr 0 \cr}\ \ \ \  {\rm and}
\ \ \ \ \pmatrix{0 \cr 1 \cr}
$$
being the set of eigenvectors with positive eigenvalues  
and
negative eigenvalues respectively. Let
\be
H_w X = X\Lambda,\label{hweig}
\ee
with
\be
X=\pmatrix{\alpha & \gamma \cr \beta & \delta \cr};\ \ \ 
\Lambda =
diag(\lambda^+_1,\cdots,\lambda^+_n,-\lambda_1^-,\cdots,-\lambda_n^-),
\label{umat}
\ee
and $\lambda_i^\pm > 0$ for all $i$.

Then,
\be
 \langle - | +\rangle= \det \alpha,\label{overform}
\ee
for the chiral determinant of left handed fermions.

Under a gauge transformation, $H_w$ will transform according to a
unitary transformation, $V$, of the form
\be
V(x,y) = e^{i\chi(x)} \delta(x-y);\ \ \ \ H_w \to H_w^g = V^\dagger H_w V.
\ee
Under this gauge transformation, 
we can set
\be
X^g=\pmatrix{V^\dagger\alpha & V^\dagger\gamma \cr V^\dagger\beta & V^\dagger\delta \cr}
\ee
as the eigenectors of $H_w^g$ with the same set of eigenvalues as $H_w$.
Since $\int \chi(x) d^2x =0$, it is clear that we have
\be
\langle-|+\rangle = \langle-|+\rangle^g 
\ee
and we have a result that is formally gauge invariant.
But this result is ambigous since the choice $X^g$ we made is not
unique. In particular, we could have an arbitrary gauge field
dependent phase that muliplies the overlap formula. 
Therefore, we have properly defined only the absolute value of the
chiral
determinant and the imaginary part of $W_+(A_\mu)$ is not defined.
The root of this problem can be seen in the continuum Pauli-Villars
analysis.
Only the real part was properly regularized in (\ref{pvreg}) and the
imaginary part was only formally regularized in the case when
anomalies
cancelled. Now, we have a formula that is finite for all gauge field
backgrounds.
The only way out for the overlap formula is to have a phase that is
ambiguous.

We can try to choose a phase in perturbation theory. If we solve for
all eigenvalues for $H_w$
when $A_\mu=0$, we can do standard perturbation theory to find
the eigenvectors in powers of $A_\mu(x)$. The problem will be that
the overlap of perturbed eigenvectors with the unperturbed ones will
have an undetermined phase. The standard approach in quantum mechanics
is to follow the Wigner-Brillouin phase choice, namely, one assumes
that
the overlap of the perturbed eigenvectors with the unperturbed ones is
real and positive. If we call $|-\rangle_0$ as the unperturbed state,
our overlap formula becomes
\be
e^{W_+^{\rm WB}(A_\mu)} = 
\frac{{}_0\langle - | -\rangle}{|{}_0\langle - | -\rangle|}
\langle -|+\rangle.
\ee
There is no reason for this to be gauge invariant and we need to see
if it the case or not. In particular, one needs to compute the
variation of $W_+^{\rm WB}(A_\mu)$ under a gauge transformation and
find out if it zero or not. If we find it to be non-zero, we still do
not have an unabiguous statement about gauge invariance.
In order to perform a careful analysis of this problem, one needs
to study the variation of $W_+(A_\mu)$ with respect to $A_\mu$
and separate the current into an ambiguous and an unabiguous piece.
This split of the current is the difference between consistent and
covariant anomaly. An important result in the overlap formalism is
the ability to find a covariant current that is unabiguous and
therefore
unabiguously identify the covariant anomaly. We will discuss
other aspects of the overlap formula before we start our discussion
of consistent and covariant anomalies. This is due to the simple fact
that the overlap formula for a vector like theory is unabiguously defined.

\subsection{Gauge field topology}

The overlap formula in (\ref{overlapl}) will be exactly zero if the
fermion number of the two ground states are not the same.
The ground state, $|+\rangle$, will
be half filled. The number of positive and negative eigenvalues of
$H_w$ need not be the same for all background gauge fields.
To see this, note that we can write $H_w$ in the form
\be
H_w = \pmatrix { B -m & C \cr C^\dagger & -B +m\cr}
\ee
where $C=\frac{1+\sigma_3}{2}\Feyn{D}$ is the chiral Dirac operator
and $B$ is the wilson term, $-\Box$.
Note that $B$ is a positive definite Hermitian matrix. If
\be
H_w \pmatrix{u\cr v\cr}=0;\ \ \  u^\dagger u + v^\dagger v=1,
\ee
then
\be
(B-m)u + Cv =0;\ \ \  C^\dagger u
-(B-m)v=0.
\ee
Therefore it follows that
\be
u^\dagger B u + v^\dagger B v = m,
\ee
which can be satisfied provided $m>0$. This also tells us that nothing
was lost in sending $\Lambda\to\infty$ in (\ref{domainwall}).

Consider
a continuous evolution from $U_\mu=1$ to $U_\mu^Q$ in (\ref{utop})
for $Q=\pm 1$.
We expect an eigenvalue of $H_w$ to cross zero at some point
in this evolution. The overlap will be exactly zero after the
eigenvalue
crosses since the ground state, $|-\rangle $, will have one less or more fermion
number
compared to $|+\rangle$. One will need to insert a fermion creation or
annihilation
operator between the two ground states to make it non-zero and this
is this expectation value of a single fermion is the zero mode in
the presence of non-zero topological charge discussed in
section~\ref{fzm}.
In essence we have a definition of the index on the lattice:
Let $n_+$ and $n_-$ be the number of positive and negative eigenvalues
of $H_w$ for a given gauge field configuration. The fermionic index is
\be
Q_f = \frac{1}{2} (n_+ - n_-).\label{index}
\ee

\subsection{Generating functional}

In order to obtain a formula for a right handed fermion
in a manner similar to (\ref{overform}) for left handed fermions, we choose
\be
M=\partial_s +m(s);\ \ \  m(s) = \cases{-m & for $s > 0 $\cr \Lambda & for
$s\le 0$\cr};\ \ \  m,\Lambda > 0.
\ee
Clearly, $M$ does not have a zero mode but $M^\dagger$ has a zero mode.
The effect of such a change is to
interchange ${\cal H}^+$ with ${\cal H}^-$.
The result is
\be
\langle + | -\rangle= \det \alpha^\dagger.
\ee
for the chiral determinant of right handed fermions.

It is important to note that we need separate copies of the many body
Hamiltonians,
one for each left handed fermion and one for each right handed fermion.
Let us therefore, refer to $a_{R,L}$ and $a^\dagger_{RL}$ as the
annihilation
and creation operators of the right and left handed fermions
respectively. Let us label the corresponding many body Hamiltonians by
${\cal H}^\pm_{R,L}$ and the ground sates by
$|\pm\rangle_{R,L}$.
Let
\be
b_{R,L}=X^\dagger a_{R,L}\label{atob}
\ee
and it follows that $b_{R,L},b_{R,L}^\dagger$ also obey canonical anti-commutation
relations.
Let us split
\be
b_{R,L}=\pmatrix{u'_{R,L}\cr d'_{R,L}\cr};
\ \ \ 
a_{R,L}=\pmatrix{u_{R,L}\cr d_{R,L}\cr},\label{baweyl}
\ee
into the $P_\pm$ pieces.
Let $|0\rangle$ be the vacuum state that is annihilated by all the
destruction
operators. Then the highest state of ${\cal H}^-_{R,L}$ is
\be
|-\rangle_{R,L} = {u'^\dagger_{R,L}}_n {u'^\dagger_{R,L}}_{n-1}\cdots 
{u'^\dagger_{R,L}}_2 {u'^\dagger_{R,L}}_1
|0\rangle
\ee
and the highest state of ${\cal H}^+_{R,L}$ is
\be
|+\rangle_{R,L} = {u^\dagger_{R,L}}_n {u^\dagger_{R,L}}_{n-1}\cdots 
{u^\dagger_{R,L}}_2
{u^\dagger_{R,L}}_1|0\rangle.
\ee

The generating functional of a left handed fermion is given by
\be  
Z_L(\bar\xi_L,\xi_L)
={}_L\langle -|e^{\xi_L d^\dagger_L +\bar\xi_L u_L}|+\rangle_L
\label{genfunl}
\ee  
and
\be 
Z_R(\bar\xi_R,\xi_R)
={}_R\langle +|e^{\xi_R u^\dagger_R +\bar\xi_R d_R}|-\rangle_R
\label{genfunr}
\ee 
for right handed fermions.
The sources, $\bar\xi_R$, $\bar\xi_L$, $\xi_R$ and $\xi_L$ are all
Grassmann variables and they anti-commute with the fermionic operators.
Note that the generating functional has the following properties.
\begin{enumerate}
\item
It does not depend
on the ordering of the operators 
since the two terms in the exponent commute with each
other
in both factors.
\item
It is clear that $d^\dagger_L$ and
$u_L$ are the propagating degrees of freedom in the first factor
since $d_L|+\rangle$ and $u^\dagger_L|+\rangle$ are both zero.
The converse holds for the second factor. 
\item
The generating functional is invariant under global chiral
transformations:
\be
\xi_R \to e^{i\varphi_R} \xi_R;\ \ 
\bar\xi_R \to \bar\xi_R e^{-i\varphi_R} ;\ \ 
\xi_L \to e^{i\varphi_L} \xi_L;\ \ 
\bar\xi_L \to \bar\xi_L e^{-i\varphi_L} .\label{chiral}
\ee
\item The propagator for right-handed and left-handed fermions are
\be
G^{ij}_L = \frac{{}_L\langle - | {d^\dagger_L}_j {u_L}_i
|+\rangle_L}{{}_L\langle-|+\rangle_L};\ \ \ 
G^{ij}_R = \frac{{}_R\langle + | {u^\dagger_R}_j {d_L}_R
|-\rangle_R}{{}_L\langle +|-\rangle_R}
\ee 
and they obey the relation
\be
G^\dagger_R = G_L.
\ee
\end{enumerate}
Since the propagators of left and right handed fermions
above are related by
hermitian conjugation
as opposed to anti-hermitian conjugation our definitions are for the
hermitian Dirac operator obtained by a multiplication of the
conventional
anti-hermitian Dirac operator by $\sigma_3$. 

\subsubsection{ Right handed fermions:} 
We will now show that 
\be
Z_R(\bar\xi_R,\xi_R)=
\left[e^{\bar\xi_R \beta\alpha^{-1} \xi_R}
\det\alpha^\dagger \right]
\label{genfunrf}
\ee
We start by noting that
we can write
\be
\bar\xi_R d_R+\xi_R u^\dagger_R  = Q_R^- + Q_R^+
\ee  
where
\be  
Q_R^+=\bar\xi_R (\delta^{-1})^\dagger d'_R +  
\xi_R u'^\dagger_R\alpha^{-1} ;\ \ \   
Q_R^-=- \bar\xi_R (\gamma\delta^{-1})^\dagger u_R
-\xi_R d^\dagger_R \beta\alpha^{-1} ,
\ee  
and we have used (\ref{umat}), (\ref{atob}) and (\ref{baweyl}). 
Since we can also write $Q_R^+$ as
\be 
Q_R^+=\bar\xi_R d_R+\xi_R u^\dagger_R  - Q_R^-
\ee 
it follows that
\be
\left[ Q_R^-, Q_R^+\right] = -\bar\xi_R \left( \beta\alpha^{-1} -
(\gamma\delta^{-1})^\dagger\right) \xi_R,
\ee 
and
\be 
e^{\bar\xi_R d_R+ u^\dagger_R \xi_R }
= e^{Q_R^-} e^{Q_R^+}  e^{\frac{1}{2} \bar\xi_R \left( \beta\alpha^{-1} -
(\gamma\delta^{-1})^\dagger\right) \xi_R}.
\ee 
We have used the identity,
\be
e^{A+B} = e^A e^B e^{-\frac{1}{2}[A,B]}
\ee
when $[A,B]$ is a c-number.
Since
\be
{}_R\langle+| e^{Q_R^-}  = {}_R\langle+|;\ \ \  
e^{Q_R^+} |-\rangle_R = |-\rangle_R,
\ee 
it follows that
\be
{}_R\langle+|e^{\bar\xi_R d_R+u^\dagger_R\xi_R }|-\rangle_R
=e^{\frac{1}{2} \bar\xi_R \left( \beta\alpha^{-1} -
(\gamma\delta^{-1})^\dagger\right) \xi_R} {}_R\langle+|-\rangle_R
= e^{\bar\xi_R \beta\alpha^{-1} \xi_R} \det\alpha^\dagger ,
\ee 
and we have used $X^\dagger X =1$ to show that 
\be
(\gamma\delta^{-1})^\dagger + \beta\alpha^{-1}=0.
\ee 

\subsubsection{ Left handed fermions:} 
We will now show that 
\be
Z_L(\bar\xi_L,\xi_L)=
\left[e^{\bar\xi_L \left[\beta\alpha^{-1}\right]^\dagger \xi_L}
\det\alpha \right].
\label{genfunlf}
\ee
In the same manner as before, we can write
\be
\xi_L d^\dagger_L +\bar\xi_L u_L = Q_L^- + Q_L^+
\ee 
where
\be 
Q_L^+=\bar\xi_L (\alpha^{-1})^\dagger u'_L + 
\xi_L d'^\dagger_L\delta^{-1} ;\ \ \  
Q_L^-=- \bar\xi_L (\beta\alpha^{-1})^\dagger d_L
-\xi_L u^\dagger_L \gamma\delta^{-1} ,
\ee 
and we have used (\ref{umat}), (\ref{atob}) and (\ref{baweyl}).
Since we can also write $Q_L^+$ as
\be
Q_L^+=\xi_L d^\dagger_L +\bar\xi_L u_L  - Q_L^-
\ee
it follows that
\be
\left[ Q_L^+, Q_L^-\right] = -\bar\xi_L \left( 
(\beta\alpha^{-1})^\dagger - \gamma\delta^{-1}\right) \xi_L,
\ee
and
\be
e^{\xi_L d^\dagger_L+ \bar\xi_L u_L }
= e^{Q_L^+} e^{Q_L^-}  e^{\frac{1}{2} \bar\xi_L \left( 
(\beta\alpha^{-1})^\dagger - \gamma\delta^{-1} \right) \xi_L}.
\ee
Since
\be
{}_L\langle -| e^{Q_L^-}  = {}_L\langle -|;\ \ \ 
e^{Q_L^-} | + \rangle_L = |+\rangle_L,
\ee
it follows that
\be
{}_L\langle -|e^{\xi_L d^\dagger + \bar\xi_L u}|+\rangle_L
=e^{ \frac{1}{2} \bar\xi_L \left( 
(\beta\alpha^{-1})^\dagger - \gamma\delta^{-1} \right) \xi_L} {}_L\langle -|+\rangle_L
= e^{\bar\xi_L (\beta\alpha^{-1} )^\dagger\xi_L} \det\alpha .
\ee

\subsection{The massless overlap Dirac operator}

A vector like gauge theory is obtained by pairing a left handed
fermion with a right handed fermion with the same charge
(same representation of the gauge group). The fermion determinant is
real and positive  since
\be
{}_R\langle-|+\rangle_R
{}_L\langle+|-\rangle_L = \det\alpha \det\alpha\dagger.
\ee
The phase choice for $|+\rangle_{R,L}$ are tied together since 
they are the ground states of identical many body operators. 
The same is true for $|-\rangle_{R,L}$. Therefore, the generating functional 
is unambiguous and does not not depend upon the 
phase choice present in the unitary matrix, $U$, that diagonalizes $H_w$.
 
In practice, one can avoid exact diagonalization of $H_w$ which is
needed for the computation of $X$ since one has
an overlap-Dirac operator for vector like
theories.
Consider the unitary operator,
\be
V = \sigma_3 \epsilon(H_w).
\ee
It follows from (\ref{hweig}) and (\ref{umat}) that
\be
\frac{1+V}{2} X = \pmatrix { \alpha & 0  \cr 0 & \delta \cr};\ \ \ 
\frac{1-V}{2} X = \pmatrix { 0 & \gamma \cr \beta & 0 \cr},\label{pmv}
\ee
and therefore
\be
\frac{1-V}{1+V} = \pmatrix{ 0 &-\left(\beta\alpha^{-1} \right)^\dagger\cr \beta\alpha^{-1} & 0
 \cr}.\label{ferprop}
\ee
Since
\be
\det X = \frac{\det\alpha}{\det\delta^\dagger};
\ee
we have the identity,
\be
\det \alpha \det\alpha^\dagger = \det\delta \det \delta^\dagger,
\ee
and therefore,
\be
\det \frac{1+V}{2} = \det\alpha\det\alpha^\dagger.\label{ferdet}
\ee
We can use (\ref{ferprop}) and (\ref{ferdet}) along with an efficient
implementation
of $V$ to compute the generating functional.
Note that the operator appearing in the determinant in (\ref{ferdet})
is not identical to the operator used to compute the propagator in
(\ref{ferprop}).

\section{Chiral anomalies}
The chiral determinant defined in (\ref{overlapl}) is a function of the
background gauge field $U_\mu(x)$. Let us combine $\mu$ and $x$
into one index, $\alpha$, and refer to the background gauge field
as $\xi_\alpha$. Variation of the $W_+(\xi)$ with respect to $\xi$
gives us the current.
Since
$|+\rangle$ does not depend on $\xi$ (we are assuming
$\Lambda\to\infty$),
the variation with respect to $\xi$ comes solely from $|-\rangle$.
Even though $H^-$ transforms covariantly under a gauge
transformation,
there is an arbitrary phase associated with $|-\rangle$
as discussed in section~\ref{chiralinf} and it need
not transform covariantly. Following standard notation, we will define
\be
j^{\rm cons}(\xi) =\partial_\alpha W_+(\xi)=
\frac{\langle\partial_\alpha -|+\rangle}{\langle -|+\rangle} d\xi_\alpha
\ee
as the consistent current. This current is an exact one form since
\be
d j^{\rm cons}=0
\ee
but it will depend upon the phase of
$|-\rangle$ and therefore ambiguous. 

It will be useful to look at the current in terms of our $\phi(x)$
(gauge invariant) and $\chi(x)$ (gauge transformation) variables.
Setting $Q=0$ and $h_\mu=0$,
\bea
j^{\rm cons} &=& 
\frac{\partial \ln \langle -|+\rangle}{\partial A_\mu(x)}
d \left[ \partial_\mu\chi(x) \right]
+\frac{\partial \ln \langle -|+\rangle}{\partial A_\mu(x)}
d \left[ \epsilon_{\nu\mu}\partial_\nu\phi(x) \right]\cr
&=&  
-\partial_\mu \left[ \frac{\partial \ln \langle -|+\rangle}{\partial A_\mu(x)}\right]
d \chi(x)
-\epsilon_{\nu\mu}\partial_\nu \left[ \frac{\partial \ln \langle
    -|+\rangle}{\partial A_\mu(x)}\right]
d \phi(x).\label{jphichi}
\eea
If the chiral determinant is gauge invariant, the first term will be
zero and the second term will only depend upon $\phi(x)$. The
difficulty arises when there is an anomaly. In this case, the first
term will not be zero. Furthermore, both terms could depend upon
$\chi(x)$. 
In order to understand the problem of anomalous gauge theories
within the context of the overlap formula, 
we will show that we can
write
\be
j^{\rm cons}(\xi) = j^{\rm cov}(\xi) + \Delta j (\xi)
\ee
where $j_\alpha^{\rm cov}(\xi)$ transforms covariantly and is
unambiguous independent of whether the underlying theory
is anomalous or not.
Even though, $\Delta j(\xi)$ will depend on the phase choice,
\be
d \Delta j = \left( \partial_\alpha \Delta j_\beta - \partial_\beta
  \Delta j_\alpha\right) d\xi_\alpha d\xi_\beta  =
{\cal F}_{\alpha\beta} d\xi_\alpha d\xi_\beta,\label{ddelj}
\ee
will be unabiguous. 

If anomalies cancel, we expect $d\Delta j =0$, implying that 
$\Delta j$ is exact. If so, we can redefine the phase of
$|-\rangle$ and get rid of $\Delta j$ making the consistent
current equal to the covariant current.
The covariant current will not have an anomaly associated with it
and the chiral determinant will be
gauge invariant.

The situation on the lattice most likely requires a fine tuning
of $H_w$ in the following sense.
Given a continuum gauge field background, we can restrict it to
a finite lattice and compute $d\Delta j$. 
For the fixed continuum gauge field, $d\Delta j$ will go to zero
as we take the continuum limit on the lattice. But, lattice spacing
effects might result in a non-zero $d\Delta j$ and there is no
phase choice that will make the chiral determinant gauge invariant
on the lattice. It is possible that a finely tuned variation of $H_w$
gives a zero $d\Delta j$. In the absence of such a lattice
Hamiltonian,
we either have to argue that the non-zero $d\Delta j$ on the lattice
will not affect continuum physics or we have to take the continuum
limit with a fixed gauge field background and then peform the
path integral over gauge fields. 

\subsection{Splitting the current}

We split $|\partial_\alpha -\rangle$ into
\be
|\partial_\alpha -\rangle = |\partial_\alpha -\rangle_\perp +
|-\rangle \langle - | \partial_\alpha -\rangle.
\ee
where
\be
\langle - |\partial_\alpha -\rangle_\perp =0
\ee
since $\langle -|-\rangle =1$.
We will show that
\be
j_\alpha^{\rm cov} = \frac{ {}_\perp\langle \partial_\alpha - | + \rangle}{\langle - |
  + \rangle},\label{covcurr}
\ee 
transforms covariantly and is unambigous.
Then,
\be
\Delta j_\alpha = \langle \partial_\alpha - | - \rangle,
\ee
is the difference between the consistent and covariant current and it
will be ambiguous.

A local gauge transformation is given by (\ref{locgauge}) 
and therefore we can write
\be
\delta\xi^g_\alpha = \delta\xi_\beta \left[{\cal D}^{-1}(g)\right]_{\beta\alpha}.
\label{dxig}
\ee
Under a gauge transformation, ${\cal H}^-$, will transform as
\be
{\cal H}^-(\xi^g) = G^\dagger(g) {\cal H}^-(\xi) G(g),\label{calhg}
\ee
where $G$ is a unitary operator on the many body space
since each $U_\mu(x)$ transforms according to (\ref{locgauge}) 
and therefore $H_w$ undergoes a unitary transformation.

Let us write,
\be
{\cal H}^-(\xi+\delta\xi) - {\cal H}^-(\xi) = 
\delta\xi_\alpha R_\alpha(\xi).\label{delcalh}
\ee
Then
\be
{\cal H}^-(\xi^g+\delta\xi^g) - {\cal H}^-(\xi^g) =  
\delta\xi^g_\alpha R_\alpha(\xi^g) =
 \delta \xi_\beta\left[{\cal D}^{-1}(g)\right]_{\beta\alpha}  R_\alpha(\xi^g),
\ee 
where we have used (\ref{dxig}).
But, using (\ref{calhg}) and (\ref{delcalh}),
\be
{\cal H}^-(\xi^g+\delta\xi^g) - {\cal H}^-(\xi^g) = 
G^\dagger(g)\left[ {\cal H}^-(\xi+\delta\xi) - {\cal H}^-(\xi) \right]
G(g)
= \delta\xi_\beta G^\dagger(g) R_\beta(\xi)
G(g).
\ee
Combining the two equations above,
\be
R_\alpha(\xi^g) 
=
\left[{\cal D}(g)\right]_{\alpha\beta} G^\dagger(g) R_\beta(\xi) G(g).
\label{ralphag}
\ee

Using standard perturbation theory,
\be
|\partial_\alpha -\rangle_\perp(\xi) = \frac{1}{{\cal H}^-(\xi) -
E_0(\xi)} \left[ \langle - | R_\alpha(\xi) |-\rangle -
R_\alpha(\xi)\right] |-\rangle;\ \ \  {\cal H}^-(\xi)|-\rangle = E_0(\xi)|-\rangle.
\ee
Under a gauge transformation,
\be
|-\rangle^g = e^{i\phi(\xi,g)} G^\dagger(g) |-\rangle,\label{groundg}
\ee
follows from (\ref{calhg}) and $\phi(\xi,g)$ is an arbitrary phase
that can depend on the background gauge field, $\xi$, and the gauge
tranformation, $g$.
Since ${\cal H}^+$ does not change under a gauge tranformation, it
follows that
\be
G(g)|+\rangle = |+\rangle.\label{ginvp}
\ee
A straight forward calculation using (\ref{ralphag}) and
(\ref{groundg}) results in
\be
|\partial_\alpha -\rangle^g_\perp(\xi) = e^{i\phi(\xi,g)} {\cal
  D}_{\alpha\beta}(g) G^\dagger(g) |\partial_\beta
-\rangle_\perp(\xi),
\ee
and therefore
\be
\left[ j^{\rm cov}_\alpha\right]^g = \left[ j^{\rm cov}_\beta\right]
\left[ {\cal D}^{-1}(g)\right]_{\beta\alpha},
\ee
using (\ref{ginvp}) showing that it transforms convariantly.
In addition, the arbitrary phase factor, $\phi(\xi,g)$, disappears and
the covariant current does not depend on the phase choice.

Following (\ref{jphichi}), we can compute the covariant anomaly by
considering
the variation with respect to $\chi$ in (\ref{covcurr}) as opposed
to $\partial_\mu \chi$.
Since the phase choice does not matter, we can follow the steps in
section~\ref{chiralinf}.
The covariant anomaly is given by
\be
{\cal A}_{\rm cov} = 1-\det \left( \alpha^\dagger V^\dagger \alpha
  +\beta^\dagger V^\dagger \beta\right),
\ee
where we have used $\det V=1$. Assuming $\chi(x)$ is infinitesimal, we
get
\be
{\cal A}_{\rm cov} = i \sum_x \chi(x) \sum_i \left(
  \alpha^\dagger_{ix} \alpha_{xi} +  \beta^\dagger_{ix} \beta_{xi}
\right) =  \frac{i}{2} \sum_x \chi(x) \tr \epsilon_{xx},\label{anomaly}
\ee
where the little trace is over the two spin components.
Since the fermionic index, $Q_f$, defined in (\ref{index}) can be
written as $\sum_x \tr \epsilon_{xx}$, the connection between anomaly
and topology follows. 

In order to show that ${\cal F}_{\alpha\beta}$ as defined in
(\ref{ddelj}) is also independent of the phase choice, we start by
noting that
\be
{\cal F}_{\alpha\beta} = 
\langle\partial_\alpha -|\partial_\beta-\rangle
- \langle\partial_\beta -|\partial_\alpha -\rangle
.\ee
Let us define the projector
\be
P=|-\rangle\langle-|;\ \ \  P^2=P,
\ee
Clearly, $P$ does not depend on the phase choice of the ground state.
A simple computation shows that
\be
{\cal F}_{\alpha\beta} = \Tr \left [ 
( \partial_\beta P ) P
 (\partial_\alpha P) - 
( \partial_\alpha P ) P
 (\partial_\beta P) \right],\label{fab}
\ee
implying that it is unabiguous.
\subsection{Sample numerical calculations}

We demonstrate the presence of ${\cal F}$ in the continuum if the
theory is anomalous and show that the cancellation is not always exact
on the lattice even if the continuum theory has no anomalies.
We also demonstrate the ambigous nature of the consistent current and
$\Delta j$ analytically by computing these quantities in lattice perturbation
theory. Finally we show the presence of topological zero modes in the
presence
of gauge fields that carry a topological change. 
We choose to perform a numerical calculation since it will help
lay out the details of an overlap calculation on the lattice.
We will set $L_1=L_2$ in what follows.
In order to not have a chirally biased presentation, we will work with
right handed fermions in the subsection.

\subsubsection{${\cal F}$ in anomalous and anomaly free cases}
We will set $\phi=\chi=0$ and also restrict ourselves to $Q=0$.
For the case of,
\be
U_\mu(n_1,n_2)= e^{i\frac{\pi h_\mu}{L}};\ \ \ \mu=1,2
\ee
the eigenfunctions of $H_w$ come in pairs of the form
\be
\frac{1}{\sqrt{2\mu(\mu-\alpha)} }\pmatrix{\beta & \mu-\alpha \cr 
  \mu-\alpha & -\beta^*\cr},\label{freeef}
\ee
with eigenvalues $\mu$ and $-\mu$ where
\bea
\mu^2&=&\alpha^2+\beta\beta^*;\cr
\alpha&=&2\sum_\mu \sin^2 \left[ \frac{\pi}{L} \left(p_\mu+\frac{h_\mu}{2}\right) \right] -
m;\cr
\beta&=&i\left\{ \sin \left[ \frac{2\pi}{L} \left(p_1+\frac{h_1}{2}\right) \right]
-i\sin \left[ \frac{2\pi}{L} \left(p_2+\frac{h_2}{2}\right) \right]\right\};\cr
&&-\frac{L}{2} \le p_\mu < \frac{L}{2}.
\eea
The chiral determinant is therefore,
\be
e^{W_-(h_\mu)} = \prod_{p_\mu} \frac{
  \beta}{\sqrt{2\mu(\mu-\alpha)} }.
\ee

Consider the continuum limit, $L\to\infty$, keeping $m>0$ fixed.
For the case of $p_\mu\approx 0$, we can write
\be
\alpha \approx 2\frac{\pi^2}{L^2}\sum_\mu \left(p_\mu+\frac{h_\mu}{2}\right)^2 - 
m;\ \ \ 
\beta \approx i\frac{2\pi}{L} \left\{ \left(p_1+\frac{h_1}{2}\right)
-i \left(p_2+\frac{h_2}{2}\right) \right\}.
\ee 
Therefore, $\mu$ leads off as $m$ and $\sqrt{2\mu(\mu-\alpha)}$ leads
off as $2m$ implying that the contribution to the chiral determinant
from these modes is proportional to $\left(p_1+\frac{h_1}{2}\right)
+i \left(p_2+\frac{h_2}{2}\right)$.
For the case of $p_1\approx \frac{L}{2}$ and $p_2\approx 0$, we can
write
\be
\alpha \approx 2\frac{\pi^2}{L^2}\sum_\mu
\left(p_\mu+\frac{h_\mu}{2}\right)^2 + 2 -
m;\ \ \ 
\beta \approx -i\frac{2\pi}{L} \left\{ \left(p_1+\frac{h_1}{2}\right)+
i \left(p_2+\frac{h_2}{2}\right) \right\}.
\ee
Let us assume that $0 < m < 2$. Then,
\be
\mu\approx (2-m)\left[ 1+ \frac{3-m}{2(2-m)^2} \frac{4\pi^2}{L^2} 
\sum_\mu
\left(p_\mu+\frac{h_\mu}{2}\right)^2\right],
\ee
and therefore, 
$\sqrt{\mu(\mu-\alpha)}$ leads off as $\frac{2\pi}{L} \sqrt{ \sum_\mu
\left(p_\mu+\frac{h_\mu}{2}\right)^2}$.
The contribution from these modes to the chiral determinant is
therefore
of order one. If $m>2$, the contribution from these modes to the
chiral
determinant would be that of a fermion with negative chirality.
In order for 
the overlap to properly reproduce a single chiral
fermion.
 The mometa
associated with the physical chiral fermion is a ball centered
around $p_1=p_2=0$ that does not touch the boundary of the
first Brillouin zone.

The phase of the eigenfunctions in (\ref{freeef}) are arbitrary.
If one used the Wigner-Brillouin phase choice, then the result is
\be
e^{W_-^{\rm WB}(h_\mu)} = e^{\frac{\pi}{8}h(h-\ub{h})}
 \frac{\vartheta(\frac{h}{2};i)}{\eta(i)};\ \ \  h=h_1+ih_2,\label{wbtheta}
\ee
in the continuum.
 We will focus on
$j_\alpha^{\rm cov}$ and ${\cal F}_{\alpha\beta}$ that do not depend on
the phase choice. We have two degrees of freedoms, namely, $h_1$
and $h_2$. 
We can compute ${\cal F}_{\alpha\beta}$ using (\ref{fab}). The
projector, $P$, has a block diagonal form with one $2\times 2$ block
for each $p_\mu$ given by
\be
P_p(h) = \frac{1}{2} -\frac{1}{2\mu}\left [ \alpha\sigma_3
  +\beta_1\sigma_1 - \beta_2\sigma_2\right] = \frac{1}{2} +
\frac{1}{2} \hat w_p(h)\cdot \vec \sigma,
\ee
where
$\beta_\mu = \frac{2\pi}{L} \left((p_\mu + \frac{h_\mu}{2}\right)$
and the components of the unit vector, $\hat w_k(h)$ are
$w_p^1(h) = \frac{\beta_1}{\mu}$,
$w_p^2(h) = -\frac{\beta_2}{\mu}$ and
$w_p^3(h) = \frac{\alpha}{\mu}$.
The only non-zero component is
\be{\cal F}_{12} (h) = \frac{i}{2} \sum_p \vec w_p \cdot \left(
\frac{\partial \vec  w_p}{\partial h_1}\wedge \frac{\partial \vec
  w_p}{\partial h_2}\label{f12toron}
\right).
\ee
In the continuum limit, $L\to\infty$, the sum over $p$ becomes an integral over a
torus and the result will become independent of $h_\mu$. The integrand
is
just the area element on the surface of a sphere therefore the result
of the integral is just a count of the number of times the map from
the
torus wraps around the sphere. This will depend on the sign on $m$ and
the result is zero wrapping for $m<0$ and one wrapping for $0<m<2$.
This is illustrated in fig.~\ref{fig1} obtained for a fixed $L$ by
peforming the sum in (\ref{f12toron}) for one arbitrary choice of
$h_\mu$. The result clearly shows the contribution from one chiral
fermion
centered around the ball at $p_\mu=0$ for $0 < m < 2$. Since the
region
around $p_1=\frac{L}{2}, p_2=0$ and $p_1=0, p_2=\frac{L}{2}$ also
contribute
for $2 < m <4$ with opposite chirality as that of the region around
$p_\mu=0$, we obtain a result corresponding to one Dirac fermion
(${\cal F}_{12}=0$) and one fermion of negative chirality. If $m>4$,
the region around $p_\mu=\frac{L}{2}$ contribute with the same
chirality as that of the region around $p_\mu=0$ resulting in a zero
result due to two Dirac fermions.

\FIGURE{
\centering
\label{fig1}
\centering
\includegraphics[scale=0.5]{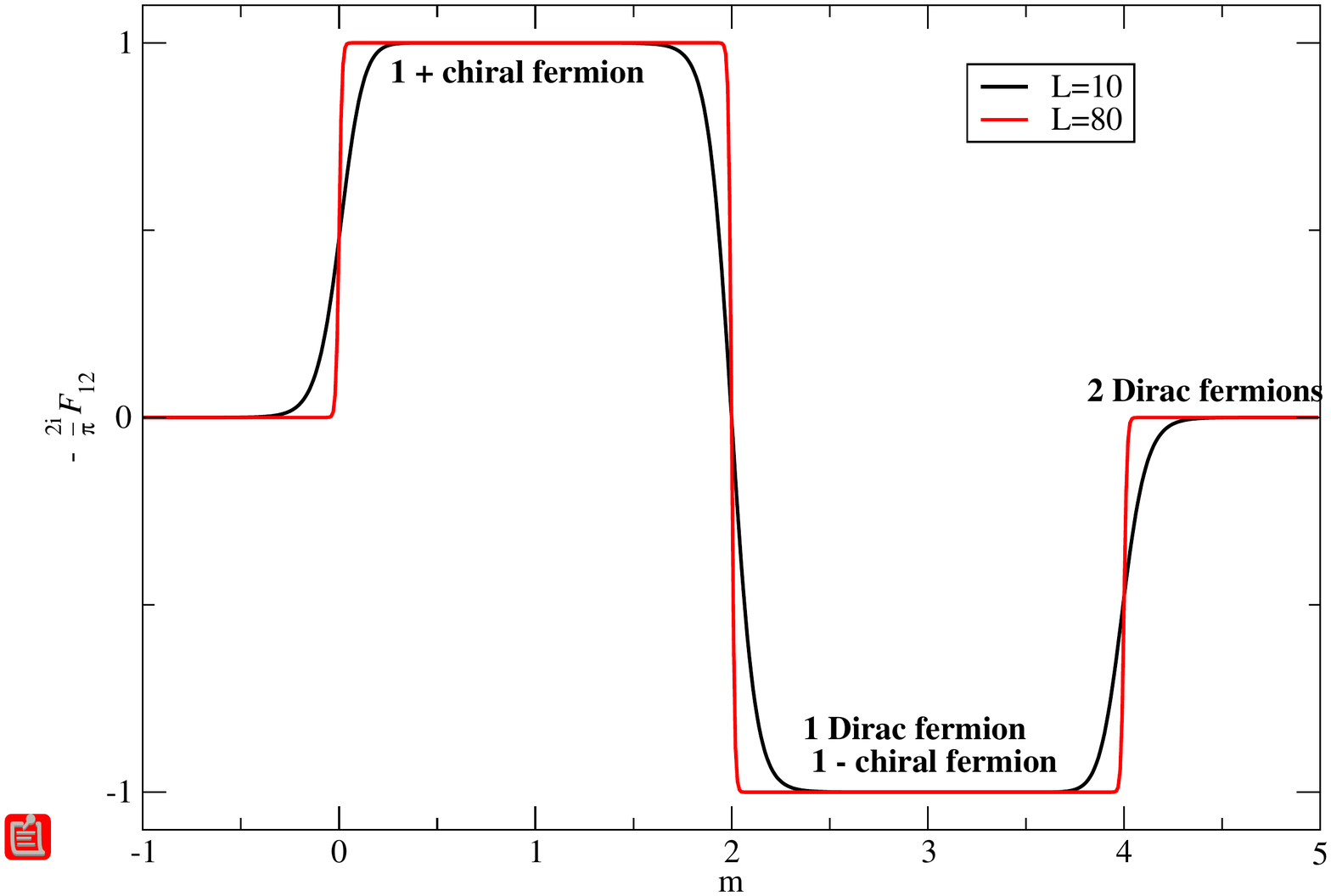}
\caption{A plot of ${\cal F}_{12}$ in (\ref{f12toron}) as a function of $m$ for $L=10,80$.}}

The result for ${\cal F}_{12}$ in the continuum limit is proportional
to $q_-^2$ and opposite in sign for fermions with negative chirality.
Therefore, we expect it to be zero for a chiral gauge theory with one
right handed fermion of charge $q_+=2$ and four right handed fermions
of
charge $q_-=1$. But, we do not expect the cancellation to be exact for
finite $L$. Therefore, even though we have chiral fermions on the
lattice we do not have exact cancellation of anomalies for a chiral
gauge theory that is anomaly free in the continuum. Furthermore, the
non-cancellation of the anomalies on the lattice is a property of the
lattice Hamiltonian alone since the computation of ${\cal F}_{12}$
only
depends on the Hamiltonian and is independent of the phase choice
of the ground states. The non-cancellation of the anomalies is related
to the dependence of ${\cal F}_{12}$ on $h_\mu$ as can be seen for
$q=2$ and
$L=8$ in fig.~\ref{fig2}. For the special case of gauge fields that
only contain $h_\mu$, we can achieve exact cancellation of anomalies
by
the following judicious choice: Let the $q_+=2$ fermion be on a $L^2$
lattice
and obey periodic boundary conditions in both directions. Let the four
$q_-=1$ fermions be on a $(L/2)^2$ lattice and obey four different
boundary conditions, namely; AA, AP, PA, PP, where A is anti-periodic
and P is periodic and the pair of symbols denote the boundary
conditions
in the two directions. Even though, each fermionic component results
in
a ${\cal F}_{12}$ that depends on $h_\mu$, the 1112 model is anomaly
free on the lattice
for this set of gauge fields as can be seen in fig.~\ref{fig2}.
This can be traced back to an identity obeyed by the theta function
appearing in (\ref{wbtheta}).

\FIGURE{
\centering
\label{fig2}
\centering
\includegraphics[scale=0.5]{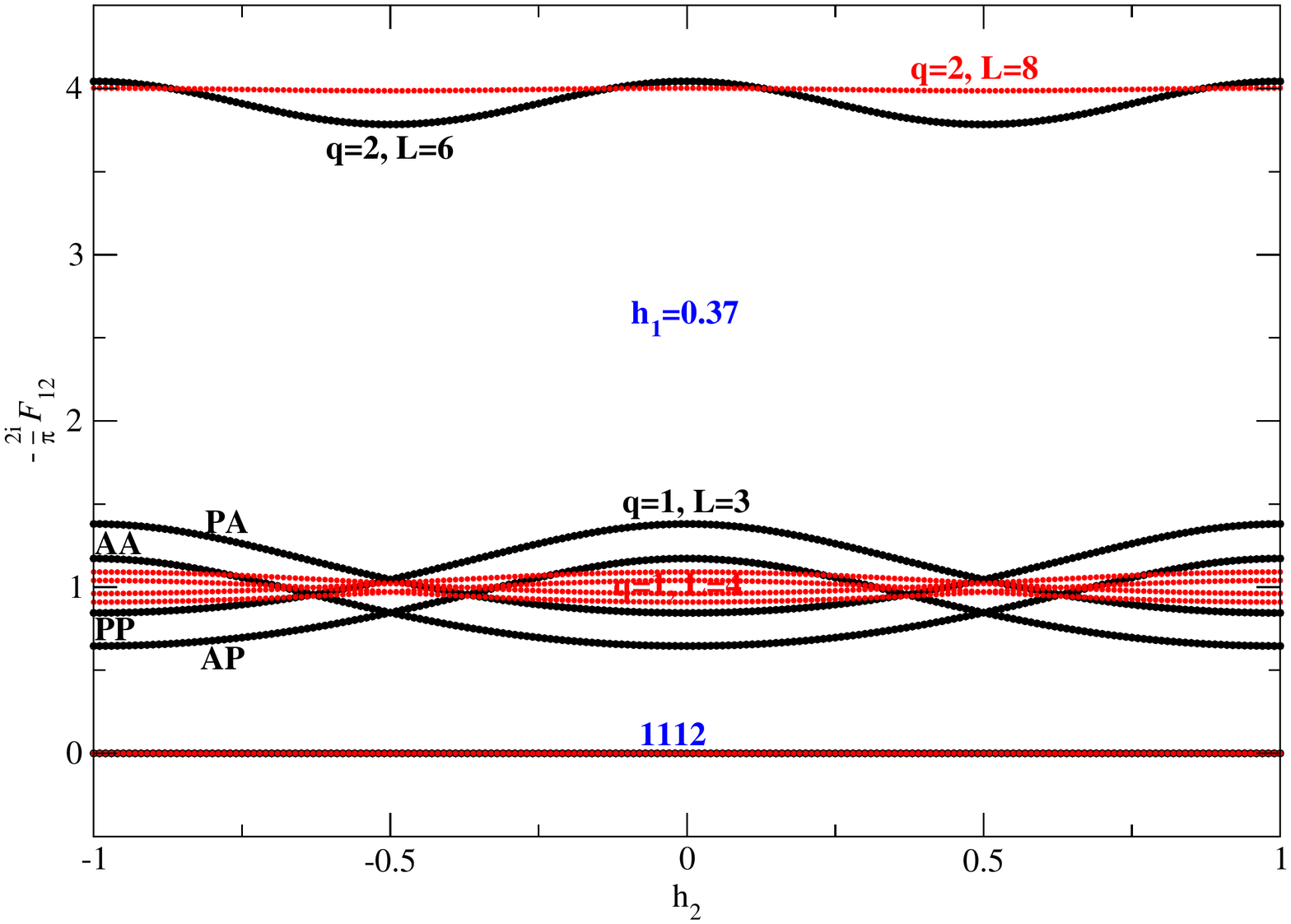}
\caption{A plot of ${\cal F}_{12}$ in (\ref{f12toron}) with $h_1=0.37$
as a function of $h_2$ for the various components of the 1112 model.}}

\subsubsection{Consistent and covariant anomalies}

As discussed in section~\ref{cpvr}, there is only a quadratic
contribution to $W_-(A_\mu)$. We may therefore write the general
expression in momentum space as
\be
W_-(A_\mu) = \sum_p
\left[ -\phi(p)\phi(-p) f(p) 
+i   \phi(p)\chi(-p) g(p) \right].\label{wfg}
\ee
We could compute $f(p)$ and $g(p)$ using lattice perturbation theory
but
our aim here is to perform a numerical computation.
We start by choosing a gauge field. 
We set $Q=0$ so that we are in
the perturbative sector. We set $h_1=h_2=\frac{1}{2}$ in
(\ref{u1u2lat})
which is equivalent to choosing anti-periodic boundary conditions.
This avoids a zero eigenvalue for the free massless Dirac operator
which would have resulted in a zero for the chiral determinant.
We set $\chi=0$ for our base gauge field configuration. We choose
$\phi$ to have a fixed momentum, namely,
\be
\phi(n_1,n_2) = \phi_0 \cos \left[ \frac{2\pi}{L} (p_1 n_1 + p_2
  n_2)\right].\label{phispec}
\ee
If we choose a value for $p_\mu$ and $\phi_0=0.1$ and set $m=1$,
we have all the ingredients to explictly write
down
$H_w$ on a finite $L\times L$ lattice. We can compute $W_-(A_\mu)$
using (\ref{overlapl}) but it would depend on the phase choice for
$|-\rangle$.
But the real part of $W_-(A_\mu)$ does not depend on the phase choice
and
a computation using several different choices for $p_\mu$ and $\phi$
will show that
\be
\lim_{L\to\infty} {\rm Re} [ W_-(\phi_0,p)] = -\frac{\pi}{2} p^2\phi_0^2.
\ee
A sample computation for $p_1=1$, $p_2=0$, $\phi_0=0.1$ as a function
of
$L$ is illustrated in Fig.~\ref{fig3}.

\FIGURE{
\centering
\label{fig3}
\centering
\includegraphics[scale=0.5]{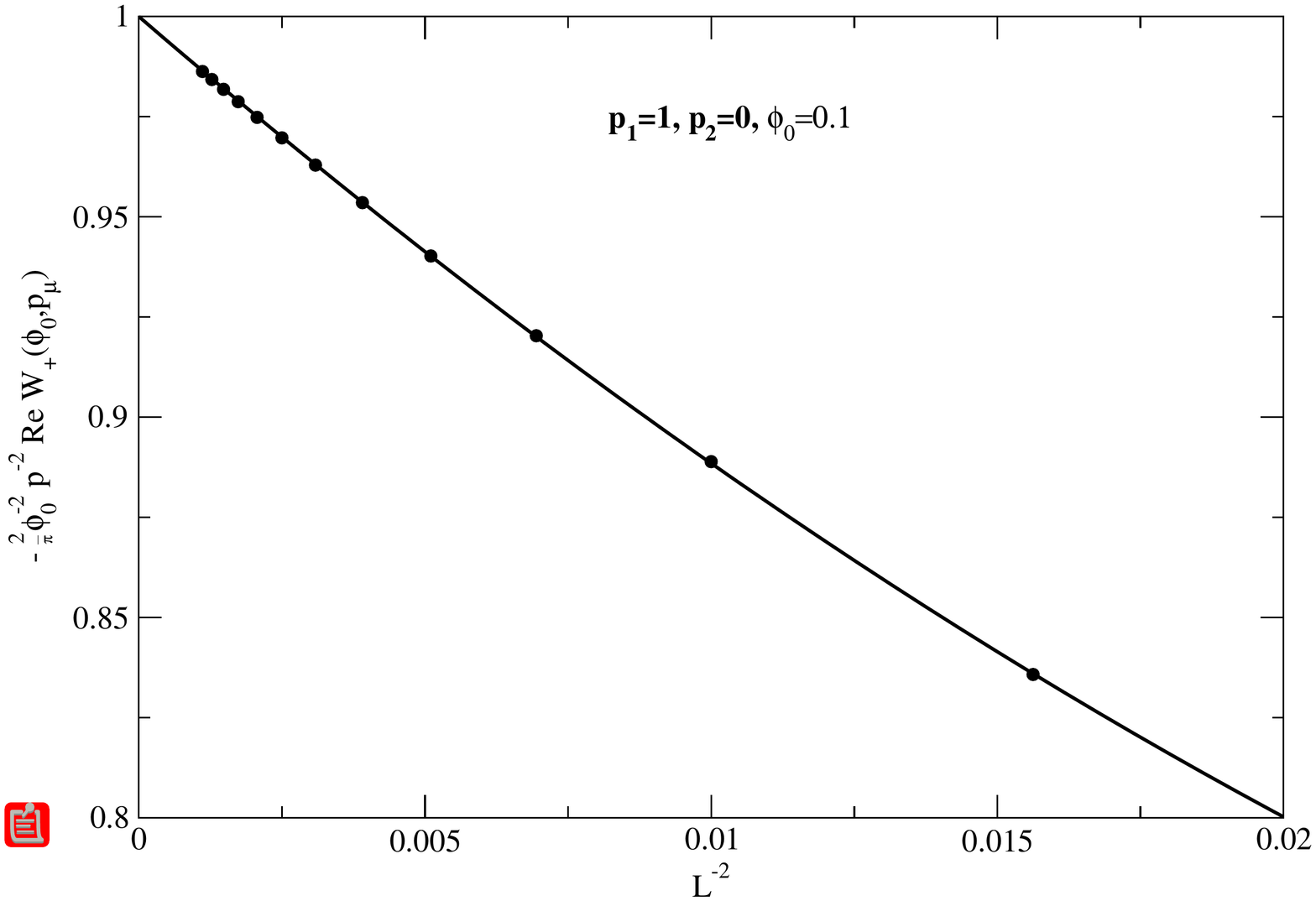}
\caption{A plot of the real part of $W_-(A_\mu)$ as a function of $L$ for a specific choice
 of
the background gauge field given by (\ref{phispec}).}}

The imaginary part of $W_-(A_\mu)$ will depend upon on the phase
choice of $|-\rangle$. One possible choice is to pick the phase of
$|-\rangle$ such
that
the imginary part is zero for all $\phi$ when $\chi=0$. Since we can
obtain the eigenstate for any $\chi$ by a gauge transformation from
$\phi=0$,
we can use this to set the phase of $|-\rangle$ for all $\chi$. 
Due to the nature of $|+\rangle$, it is easy to see that the imaginary
part will be zero for all $\phi$ and $\phi$. Another possible choice
is the Wigner-Brillouin phase choice discussed in section~\ref{chiralinf}.
This will not result in the imaginary part of
$W_-^{\rm WB}(A_\mu)$
being zero for all $A_\mu$. But, we can compute the consistent and
covariant
anomalies by computing these currents by varying $\chi$ for a fixed
$\phi$ as discussed in (\ref{jphichi}).
We choose $\chi$ to have the same momentum as $\phi$ to obtain
a non-zero anomaly. The results are shown in Fig.~\ref{fig4} and
Fig.~\ref{fig5}.
The covariant anomaly is non-zero and independent of the phase
choice as expected but the consistent anomaly depends on the phase
choice. The presence of a covariant anomaly unabiguously shows that the continuum
theory is anomalous. One can proceed like in the previous section and
study
the 1112 model. For smooth backgrounds like the one considered here,
the model will look anomaly free on the lattice. But, this will not be
the case for an abitrary choice of $\phi$ and $\chi$. The special
choice
of boundary conditions and lattices prescribed in the previous
subsection will not render the lattice theory anomaly free for an
arbitrary
$\phi$ and $\chi$. 

\FIGURE{
\centering
\label{fig4}
\centering
\includegraphics[scale=0.5]{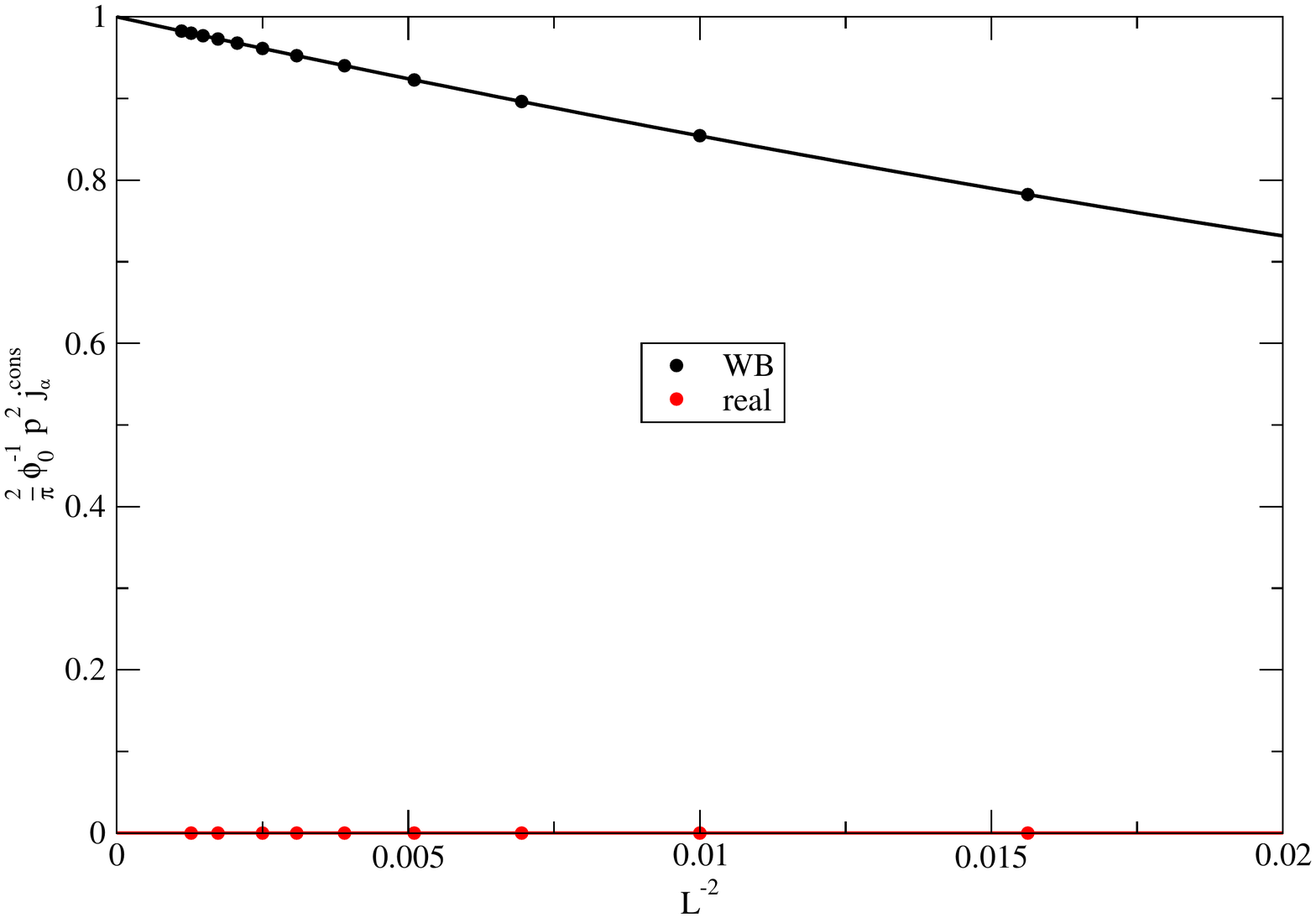}
\caption{A plot of the consistent anomaly as a function of $L$ for a specific choice
of
the background gauge field given by (\ref{phispec}).}}

\FIGURE{
\centering
\label{fig5}
\centering
\includegraphics[scale=0.5]{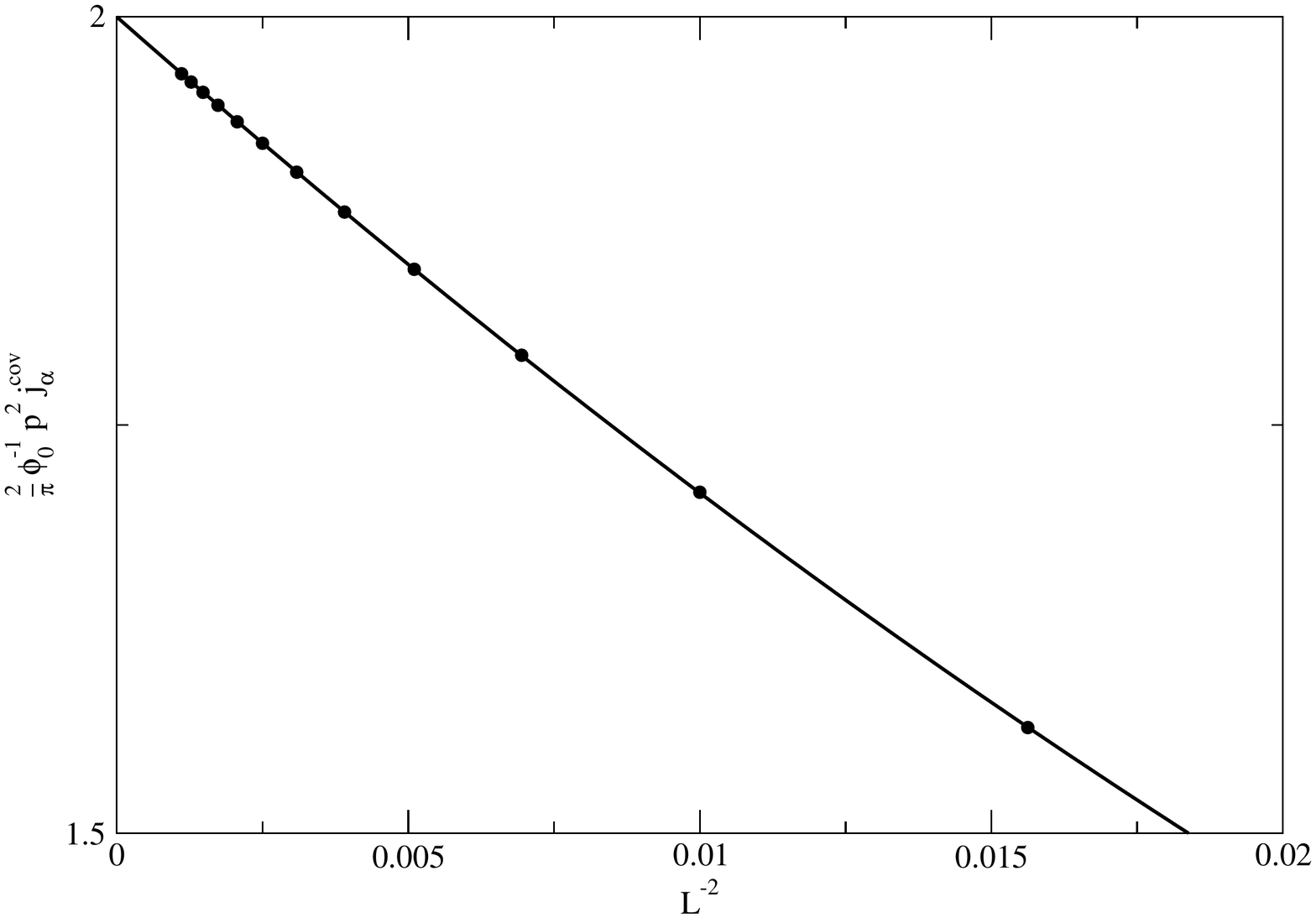}
\caption{A plot of the covariant anomaly as a function of $L$ for a specific choice
of
the background gauge field given by (\ref{phispec}).}}

\subsubsection{Topology on the lattice}

The gauge field configuration given in (\ref{utop}) has a topological
charge equal to $Q$ for integer values of $Q$. But, the configuration
is well defined on the lattice for any real value of $Q$ and {\sl
  smoothly}
interpolates between different topological sectors. This is clearly
a notion that is not present in the continuum since (\ref{winding})
can
only be defined for integer values of $Q$. Furthermore, we also know
that
$Q$ has to be an integer in (\ref{qtop}). 
We can compute $Q_f$ as defined in (\ref{index}) for a choice of $Q$.
We fix $Q=1$ and plot $Q_f$ as a function of $m$ for two different
values
of $L$ in fig.~\ref{fig6}. We make two remarks. The index does not switch
from $Q_f=0$ to $Q_f=1$ exactly at $m=0$ for $L=5$. This is a lattice
artifact which goes away in the continuum limit. The index for $2 < m
< 4$ is that of a fermion with negative chirality since we have one
Dirac
fermion and one fermion of negative chirality in this region. The
index is
zero for $m>4$ since we have two dirac fermions.
Treating $Q$ as a real number in (\ref{utop}), we plot $Q_f$ vs $Q$
for $L=8$ and $m=1$ in fig.~\ref{fig7}. Clearly, $Q_f$ has to be an
integer and the switch occurs for a $Q$ close to half integer. The
lattice field configuration close to switch does not have a continuum
anolog.

\FIGURE{
\centering
\label{fig6}
\centering
\includegraphics[scale=0.5]{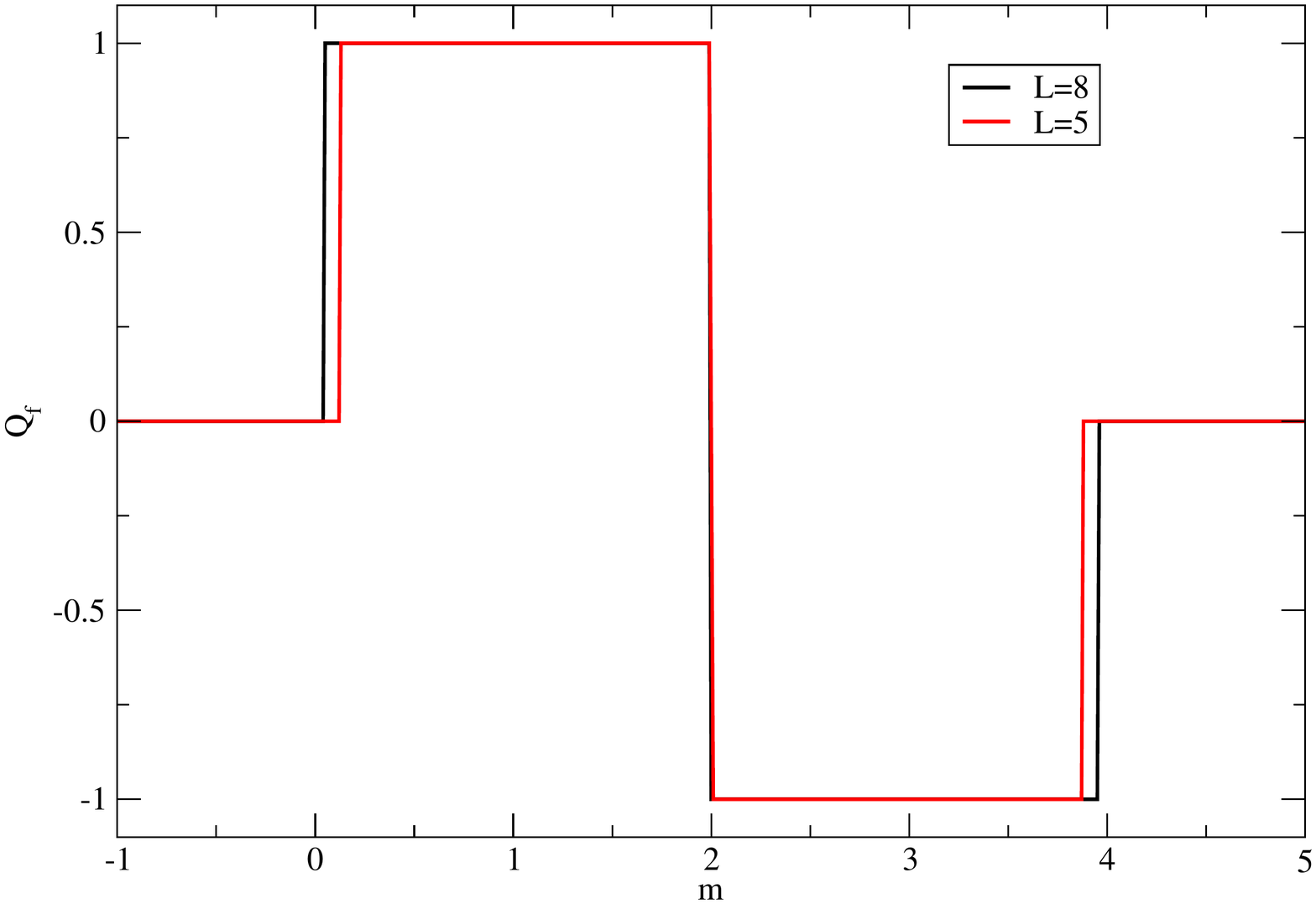}
\caption{A plot of the fermionic index as a function of $m$ for $L=5,8$.}}

\FIGURE{
\centering
\label{fig7}
\centering
\includegraphics[scale=0.5]{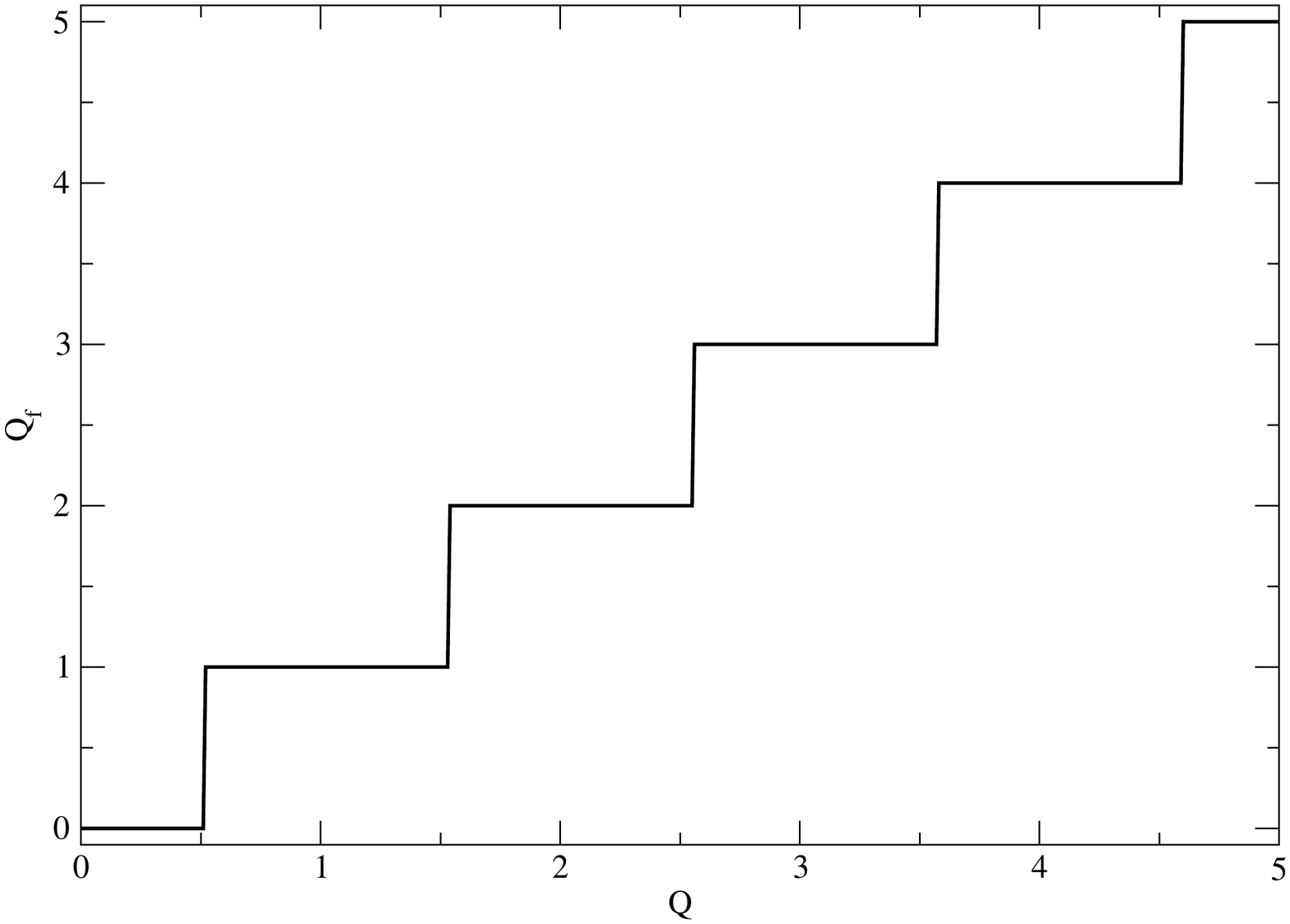}
\caption{A plot of the fermionic index as a function of $Q$ for $L=8$
 and $m=1$.}}

We can take the uniform $Q=-1$ background on
a $L^2$ lattice and compute $\tr \epsilon_{xx}$ to obtain the anomaly as
per (\ref{anomaly}). We will find this to be uniform in $x$ and equal
to $-\frac{1}{L^2}$ such that the sum over the lattice gives the
correct value of $Q$. We can add a $\phi$ component to the uniform
background
and we can compute the local distribution of topological charge. 
There is a single zero mode in the continuum given by
(\ref{zeromodes}) when $Q=-1$. 
We can compute the unnormalized zero mode on the lattice by computing
\be
\psi(x) = {}_L\langle - | u_L (x) |+\rangle_{L}
\ee
as per the generating functional in (\ref{genfunl}) where $u_L(x)$ stands for the
operator at the location $x$. We plot $\frac{\psi(x,x)}{\psi(0,0)}$,
the diagonal part of the zero mode, in
fig.~\ref{fig8}
and compare it with the corresponding function in the continuum.

\FIGURE{
\centering
\label{fig8}
\centering
\includegraphics[scale=0.5]{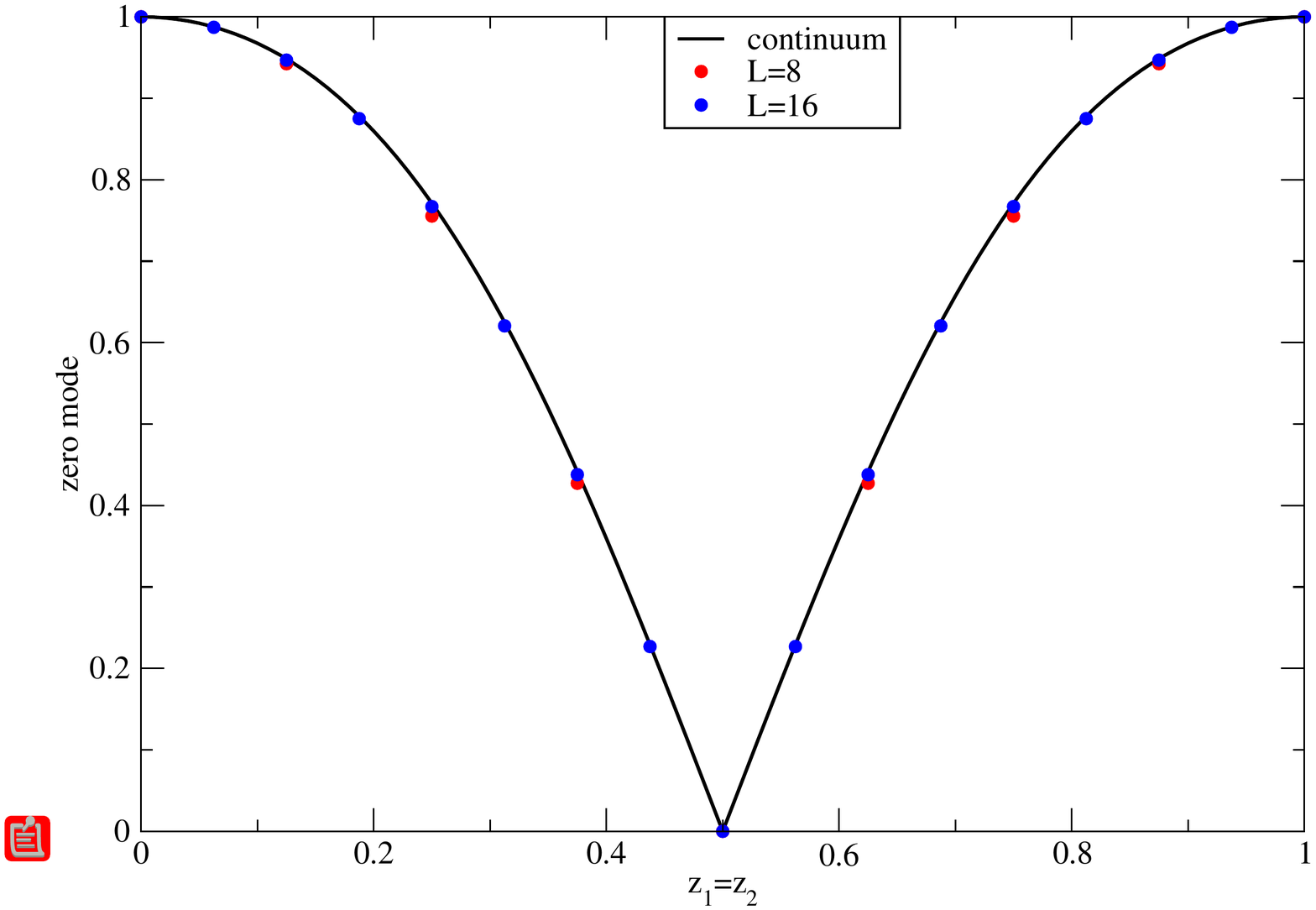}
\caption{A plot of the fermionic zero mode restricted to the diagonal
  on the torus.}}

\section*{Acknowledgements}
I would like to thank Sourendu Gupta, Saumen Datta, Nilmani Mathur and
Rajiv Gavai for organizing a wonderful school. Participating students
provided
a stimulating atmosphere for the lecturers and it was a pleasure to
discuss various related physics issues with the participants during
the school. I would like to specially thank my
Japanese
colleagues who participated in this school in spite of the major
earthquake
that hit Japan on March 11 and seriously affected the country as a
whole.
I am deeply indebted to my senior colleague, Herbert Neuberger, with
whom I have collaborated productively for several years.
The author acknowledges partial support by the NSF under grant number
PHY-0854744.

\section*{Bibliography}
There are several books that discuss chiral gauge theories. One
example is
\hfill\break
S.~Weinberg,
  ``The quantum theory of fields. Vol. 2: Modern applications,''
  Cambridge, UK: Univ. Pr. (1996) 489 p.
\hfill\break
Chapter 5 in
\hfill\break
G.~'t Hooft,
  ``Under the spell of the gauge principle,''
  Adv.\ Ser.\ Math.\ Phys.\  {\bf 19}, 1-683 (1994).
\hfill\break
contains three important papers on the physical effects of chiral symmetry and gauge field topology.
\hfill\break
There are several reviews on chiral gauge theories. Some relevant ones are
\hfill\break
M.~E.~Peskin,
  ``Chiral Symmetry And Chiral Symmetry Breaking, Les Houches
  Sum.School 1982:217".
\hfill\break
R.~D.~Ball,
  ``Chiral Gauge Theory,''
  Phys.\ Rept.\  {\bf 182}, 1 (1989).
\hfill\break 
H.~Neuberger,
  ``Summary talk at CHIRAL '99,''
  Chin.\ J.\ Phys.\  {\bf 38}, 735-743 (2000).
  [hep-lat/9912020].
\hfill\break 
H.~Neuberger,
  ``Chiral symmetry outside perturbation theory,''  
  [hep-lat/9912013].
\hfill\break 
E.~Poppitz, Y.~Shang,
  ``Chiral Lattice Gauge Theories Via Mirror-Fermion Decoupling: A Mission (im)Possible?,''
  Int.\ J.\ Mod.\ Phys.\  {\bf A25}, 2761-2813 (2010).
  [arXiv:1003.5896 [hep-lat]].
\hfill\break
Two dimensional abelian gauge theory coupled to a single massless
Dirac fermion was solved by Schwinger in
\hfill\break
 J.~S.~Schwinger,
  ``Gauge Invariance and Mass. 2.,''
  Phys.\ Rev.\  {\bf 128}, 2425-2429 (1962).
\hfill\break
and is referred to as the Schwinger model.
\hfill\break
A detailed discussion of the Schwinger model including an analysis of
the zero modes in the non-zero topological sector can be found in
\hfill\break
I.~Sachs, A.~Wipf,
  ``Finite Temperature Schwinger Model,''
  Helv.\ Phys.\ Acta {\bf 65}, 652-678 (1992).
  [arXiv:1005.1822 [hep-th]].
\hfill\break
There are several text books on lattice gauge theories that discuss
fermion doublers and Wilson fermions. One example is
\hfill\break
 M.~Creutz,
  ``Quarks, Gluons And Lattices,''
  Cambridge, Uk: Univ. Pr. ( 1983) 169 P. (Cambridge Monographs On
  Mathematical Physics).
\hfill\break
This has been very recently supplemented by the review article,
\hfill\break
M.~Creutz,
  ``Confinement, chiral symmetry, and the lattice,''
  [arXiv:1103.3304 [hep-lat]].
\hfill\break
Pauli-Villars regularization of anomaly free gauge theories was first
dicussed in
\hfill\break
 S.~A.~Frolov, A.~A.~Slavnov,
  ``An Invariant regularization of the Standard Model,''
  Phys.\ Lett.\  {\bf B309}, 344-350 (1993).
\hfill\break
A proposal to regulate chiral fermions on the lattice using a mass
defect was developed by
\hfill\break
D.~B.~Kaplan,
  ``A Method for simulating chiral fermions on the lattice,''
  Phys.\ Lett.\  {\bf B288}, 342-347 (1992).
  [hep-lat/9206013].
\hfill\break
The above two independent ideas to regulate chiral gauge theories was
shown to be equivalent in
\hfill\break
R.~Narayanan, H.~Neuberger,
  ``Infinitely many regulator fields for chiral fermions,''
  Phys.\ Lett.\  {\bf B302}, 62-69 (1993).
  [hep-lat/9212019].
\hfill\break 
The overlap formula was derived starting from the path integral
formalism of fermions coupled to a mass defect by using the second
quantized picture in
\hfill\break 
R.~Narayanan, H.~Neuberger,
  ``Chiral determinant as an overlap of two vacua,''
  Nucl.\ Phys.\  {\bf B412}, 574-606 (1994).
  [hep-lat/9307006].
\hfill\break
This paper also discusses the Wigner-Brillouin phase choice and
computes the anomaly. In addition, it also explains how gauge field
topology
on the lattice is related to the flow of eigenvalues of the Hermitian
Wilson-Dirac operator.
\hfill\break 
A specific example illustrating the connection between topology and
eigenvalues of the Hermitian Wison-Dirac operator can be found in
\hfill\break
 R.~Narayanan, H.~Neuberger,
  ``Chiral fermions on the lattice,''
  Phys.\ Rev.\ Lett.\  {\bf 71}, 3251-3254 (1993).
  [hep-lat/9308011].
\hfill\break
A detailed discussion of the overlap formalism and several examples
illustrating computations of the anomaly and topological charge can be
found in
\hfill\break
R.~Narayanan, H.~Neuberger,
  ``A Construction of lattice chiral gauge theories,''
  Nucl.\ Phys.\  {\bf B443}, 305-385 (1995).
  [hep-th/9411108].
\hfill\break 
An explicit expression for the overlap Dirac operator in a vector like
gauge theory was first written down in
\hfill\break
 H.~Neuberger,
 ``Exactly massless quarks on the lattice,''
  Phys.\ Lett.\  {\bf B417}, 141-144 (1998).
  [hep-lat/9707022].
\hfill\break 
Consistent and covariant currents originally appear in
\hfill\break
  W.~A.~Bardeen, B.~Zumino,
  ``Consistent and Covariant Anomalies in Gauge and Gravitational Theories,''
  Nucl.\ Phys.\  {\bf B244}, 421 (1984).
\hfill\break 
The consistent and covariant anomalies were correctly identified for
overlap fermions in
\hfill\break
  S.~Randjbar-Daemi, J.~A.~Strathdee,
  ``Consistent and covariant anomalies in the overlap formulation of chiral gauge theories,''
  Phys.\ Lett.\  {\bf B402}, 134-140 (1997).
  [hep-th/9703092].
\hfill\break 
A complete understanding of the covariant anomaly and the Berry's
curvature for overlap fermions along with two non-perturbative
examples can be found in
\hfill\break
  H.~Neuberger,
  ``Geometrical aspects of chiral anomalies in the overlap,''
  Phys.\ Rev.\  {\bf D59}, 085006 (1999).
  [hep-lat/9802033].
\hfill\break
The above set of references serve only as a guide to further
understand the topics covered in these lectures. The lectures
themselves
did not cover all developments pertaining to overlap fermions.
Papers cited by the above list of references will provide a further
understanding of lattice chiral gauge theories and citations of
the above references lead to further developments.

\end{document}